\documentclass[aps,pre,groupedaddress]{revtex4}%PRE
\usepackage{amsmath,amssymb,amsthm,mathrsfs,amsfonts,dsfont} 
\usepackage{graphicx}
\usepackage{subfigure}
\usepackage{amsbsy}
\usepackage{bm}
\usepackage{xcolor}
\usepackage{dcolumn}
\newcommand{\blue}[1]{{\color{black}#1}}
\usepackage{algorithm}
\usepackage{algpseudocode}
\usepackage{algorithmicx}

\usepackage{amsmath}
\usepackage{array}
\usepackage{booktabs}
\usepackage{physics}

\def\id{\mathds{1}}
\def\J{\mathbf{J}}
\newcommand{\matr}[1]{\mathbf{#1}} 

\begin{document}

\title{Neural Langevin Machine: a local asymmetric learning rule can be creative}
\author{Zhendong Yu$^{1}$}
\thanks{Equal contribution.}
\author{Weizhong Huang$^{1}$}
\thanks{Equal contribution.}
\author{Haiping Huang$^{1,2}$}
\email{huanghp7@mail.sysu.edu.cn}
\affiliation{$^{1}$PMI Lab, School of Physics,
Sun Yat-sen University, Guangzhou 510275, People's Republic of China}
\affiliation{$^{2}$Guangdong Provincial Key Laboratory of Magnetoelectric Physics and Devices,
Sun Yat-sen University, Guangzhou 510275, People's Republic of China}
\date{\today}

\begin{abstract}
	Fixed points of recurrent neural networks can be leveraged to store and generate information. These fixed points are captured by the Boltzmann-Gibbs measure, which leads to 
	neural Langevin dynamics that relax to those fixed points for generative learning of a real dataset. We call this type of generative model a neural Langevin machine, which derives an asymmetric and firing-rate-speed-adjusted learning rule requiring only local neural signals, thereby bearing biological relevance in terms of local predictive learning. An out-of-equilibrium regime of the generative process is revealed, together with a memorization-to-generalization transition with increasing training data size. The neuro-inspired machine can also realize a continuous exploration of the phase space for different kinds of generative images and can denoise a corrupted image as well. 
	\end{abstract}

 \maketitle

%%%%%%%%%%%%%%%%%%%%%%%%%%%%%%%%%%%%%%%%%%%%%%%%%%%%%%%%%%%%%%%%%
\section{Introduction}
Generative models play an important role in modern machine learning, because of their wide applications in generating data samples after a complex non-analytic distribution is learned~\cite{Hinton-2002,VAE-2019,NEDM-2015,GAN-2014}. These data samples cover a diverse range of high-dimensional vectors such as natural images, language texts, and even neural activities in the brain. Earlier representative energy-based generative models include restricted Boltzmann machine~\cite{Hinton-2002,Huang-2020}, and recent neural-networks-based developments include variational auto-encoder~\cite{VAE-2019}, generative adversarial networks~\cite{GAN-2014}, and a currently-dominant generative diffusion models~\cite{NEDM-2015,Song-2020,Yu-2025}. In generative diffusion models, the forward stochastic dynamics and the backward score-driven stochastic dynamics are implemented~\cite{Yu-2025}. The score is actually the gradient of the log-data-likelihood in data space, approximated by a complex neural network that must be trained using the forward trajectories.

The current artificial-neural-network-based generative models are criticized for lacking biological plausibility and are complex in training, requiring a bag of tricks. First, biological neural networks bear a recurrent structure where synaptic couplings (and plasticity) are asymmetric from one neuron to another~\cite{Maass-2009,ND-2014,Plast-2020}. Currently, popular generative models miss this brain-like feature. Second, the neural dynamics are used as a sampling machine, such as a probabilistic computation via neural response 
variability~\cite{MCNN-2002,WS-2023,PCsamp-2024}, and fixed points or chaotic attractors can be leveraged to store and generate examples. The  neural response variability is an important feature of brain dynamics or neural computation. Third, the learning is always local in a biological circuit and thus efficient, while most generative models requires a global artificial error signal~\cite{Hinton-2020}. Recent works have started to consider the role of chaotic activity in generating data samples~\cite{PU-2025}. However, the learning rule is still of the symmetric Hebbian type, despite a random, untrained asymmetric baseline being considered. Another recent work proposed a Langevin dynamics with an Ising-type potential~\cite{Potential-2025}, but the learning is driven to reverse the forward dynamics in the most thermodynamically reversible way, which violates the above three requirements due to biological restrictions and thus does not pave the way towards understanding perception and imagination in the brain~\cite{RMP-2006,Huang-2024}.

To take a trade-off between biological relevance and engineering purpose, we propose a neural Langevin machine (NLM) that satisfies the above three criteria, i.e., local learning, neural-dynamics-based sampling, and asymmetric recurrence. This NLM has the benefit of shedding light on neuro-inspired generative models. In particular, the generating process yields transitions among different types of data samples, resembling continuous brain dynamics driven by background stochastic noise (ubiquitous in neural circuits~\cite{Neuron-2016,Csamp-2020}). Hence, our work provides a promising avenue for a neural-dynamics-based generative model that allows for a dynamic exploration of the creative activity space of imagination, e.g., transition from memory to generalization. In addition, the local learning rule can be interpreted as (firing rate) speed-adjusted predictive learning, being of particular biological plausibility~\cite{Xie-1999,PC-1999,PCsamp-2024,LE-2021}. This rule is derived from a minimal entropy principle about the kinetic energy landscape of recurrent neural dynamics, connecting generative models to fundamental physical principles ~\cite{Qiu-2025,Wang-2024,Du-2024,HWZ-2025}. Remarkably, the sampling process coincides with the hypothesis of Bayesian brain without probability~\cite{BB-2016}, while the energy principle is consistent with the new path towards artificial intelligence proposed by Yann LeCun~\cite{Lecun-2024}.

\section{Neural Langevin machine as a generative model}\label{prob}
The dynamics of recurrent neural networks with random asymmetric coupling among neurons have a chaos transition when the synaptic gain crosses a threshold~\cite{Chaos-1988,Zou-2024}. After the chaos transition, the emergence of exponentially many unstable fixed points supports the chaos~\cite{Helias-2025}. It remains unknown if and how these fixed points can be used for learning (\blue{note that learning may alter the topological structure of fixed-points}). To investigate this point, we employ an alternative neural dynamics~\cite{Qiu-2025}:
\begin{equation}
    \frac{d \mathbf{x}}{d t} = -\nabla_{\mathbf{x}} E(\mathbf{x}) + \sqrt{2T} \boldsymbol{\epsilon}(t), \label{eq:langevin}
\end{equation}
where $\mathbf{x} \in \mathbb{R}^N$ is the synaptic current vector. The potential (kinetic) energy function $E(\mathbf{x})$ is defined as:
\begin{equation}
    E(\mathbf{x}) = \frac{1}{2}\sum_{i=1}^N \left( -x_i +b_i+ \sum_{j=1}^N J_{ij} \phi(x_j) \right)^2. \label{eq:energy_func}
\end{equation}
In this paper, we set the firing rate function $\phi (x) = \tanh (x)$. $b_i$ indicates bias. Here $T$ represents a temperature parameter, and $\boldsymbol{\epsilon}(t)$ is an $N$-dimensional Gaussian white noise vector, characterized by zero mean $\langle \epsilon_i(t) \rangle = 0$ and temporal correlation $\langle \epsilon_i(t) \epsilon_j(t') \rangle = \delta_{ij} \delta(t-t')$. $J_{ij}$ denotes the neural coupling from neuron $j$ to $i$. They are always asymmetric, thereby allowing for no potential function for the original chaotic neural dynamics~\cite{Chaos-1988}.

 The $i$-th component of the gradient force $-\nabla_{\mathbf{x}} E(\mathbf{x})$, denoted as $F_i$, is given by:
\begin{equation}
    F_i \equiv -\frac{\partial E(\mathbf{x})}{\partial x_i} = -x_i + h_i - \phi'(x_i)\sum_{j=1}^N J_{ji} (h_j - x_j), \label{eq:force_comp}
\end{equation}
where we introduce the auxiliary variable $h_i$ as integrated synaptic current $h_i \equiv \sum_{j=1}^N J_{ij} \phi(x_j)+b_i$. The first two terms are exactly the same driving force in the classic random neural network~\cite{Chaos-1988}. The last term in the right-hand-side of Eq.~\eqref{eq:force_comp} is the very Onsager feedback term, which plays an important role in stabilizing the chaotic fluctuations~\cite{HWZ-2025}, making those (exponentially many) unstable fixed points usable! \blue{The dynamics we study here share the same set of fixed-points with those studied in the classic random neural network given random couplings defined above~\cite{Chaos-1988}. In fact, after a generative learning, the synaptic couplings are not random, and thus the fixed points are not the same as in the random networks. We shall expand this point in the results section.} \blue{This dynamics formula could be potentially connected to a structured pyramidal neuron with bottom-up input via its basal dendrites (the first two terms in $F_i$) and top-down input via its apical dendrites (the last term in $F_i$)~\cite{LE-2021}.}

\blue{We emphasize that our kinetic energy perspective shares the same spirit with the latent equilibrium~\cite{LE-2021} or equilibrium propagation~\cite{EP-2017}. The former derived the neuronal dynamics on a constant-energy manifold with an abstract network sate combining network activity and its velocity, rather than gradient descent in the kinetic energy surface in our perspective, while the latter proposed a total energy as a linear combination of internal energy and external cost function and thus nudged and free fixed points are used to update the model parameters, in contrast to our unsupervised learning case where the speed-adjusted contrastive learning is used.}

The system described by Eq. \eqref{eq:langevin} converges to an equilibrium distribution, $P(\mathbf{x})$, which takes the form of a Boltzmann-Gibbs distribution~\cite{Risken-1996}:
\begin{equation}
    P(\mathbf{x}) = \frac{1}{Z} \exp(-\beta E(\mathbf{x})), \label{eq:boltzmann_dist}
\end{equation}
where $\beta = 1/T$ is the inverse temperature, and $Z = \int \exp(-\beta E(\mathbf{x})) d\mathbf{x}$ is the partition function, ensuring normalization. Therefore, the recurrent dynamics [Eq.~\eqref{eq:langevin}] in the zero-temperature limit has the same set of fixed points as the classic random neural network where the driving force  $F_i = -x_i + h_i$~\cite{Chaos-1988}, from which the kinetic energy $E=\frac{1}{2}\sum_iF_i^2$ follows, as shown in Eq.~\eqref{eq:energy_func}.  By expanding $E(\mathbf{x})$, we get the following equivalent form:
\begin{equation}\label{Hami}
E(\mathbf{x})=\frac{1}{2}\sum_ix_i^2-\sum_{i,j}J_{ij}x_i\phi_j+\frac{1}{2}\sum_{j,k}\mathcal{K}_{jk}\phi_j\phi_k,
\end{equation}
where $\phi_i=\phi(x_i)$, $\mathcal{K}_{jk}=\sum_iJ_{ij}J_{ik}$. Bias can be included as an extra degree $J_{i,N+1}=b_i$ and $\phi_{N+1}=1$. Therefore, $\mathbf{b}$ will not be separately mentioned below. This Hamiltonian form is very different from the Ising model form used in Boltzmann machine~\cite{Hinton-2002} or the recent thermodynamic computing framework mentioned before~\cite{Potential-2025}. The last two terms in Eq.~\eqref{Hami} contribute to the stabilization of the original unstable fixed points in the classic random neural network, making the generative function possible.

Our objective is then to optimize the coupling parameters $\{J_{kl}\}$ such that the model distribution $P(\mathbf{x})$ closely approximates a given data distribution $P_{data}(\mathbf{x})$. The discrepancy between these two distributions is quantified by the Kullback-Leibler (KL) divergence:
\begin{equation}
\begin{aligned}
   & \mathcal{L}_{KL}(P_{data} || P) = \int P_{data}(\mathbf{x}) \ln \frac{P_{data}(\mathbf{x})}{P(\mathbf{x})} d\mathbf{x} \\
       &= \langle \ln P_{data}(\mathbf{x}) \rangle_{data} + \ln Z + \beta \langle E(\mathbf{x}) \rangle_{data}. \label{eq:kl_expanded}
\end{aligned}
\end{equation}
Here, $\langle \cdot \rangle_{data}$ denotes an expectation taken with respect to $P_{data}(\mathbf{x})$. Since the term $\langle \ln P_{data}(\mathbf{x}) \rangle_{data}$ is independent of the model parameters, minimizing $\mathcal{L}_{KL}$ is equivalent to minimizing the objective function $\mathcal{L}(\J) = \ln Z + \beta \langle E(\mathbf{x}) \rangle_{data}$. If we define $\ln Z$ as a negative free energy, this objective function bears the physical meaning of entropy in physics.
This minimum entropy principle leads to the following learning rule:
\begin{equation}\label{eq:kl_grad_final}
%\begin{aligned}
    \frac{\partial \mathcal{L}}{\partial J_{kl}} 
    = \beta \left[ \left\langle (x_k - h_k) \phi(x_l) \right\rangle_{model} - \left\langle (x_k - h_k) \phi(x_l) \right\rangle_{data} \right]. 
%\end{aligned}
\end{equation}
 A Detailed derivation is given in the Appendix~\ref{SM-a}. 
 
Although this rule bears a similar form as contrastive divergence used in restricted Boltzmann machine~\cite{Hinton-2002},  two key differences are clear. First, this rule is not of the simple symmetric Hebbian type, but is asymmetric in the sense of neuron index permutation. Second, the term $x_k-h_k$ represents the postsynaptic contribution and bears the physical meaning of prediction [reducing the speed of the postsynaptic neuron activity, see Eq.~\eqref{eq:energy_func}],
which is surprisingly consistent with predictive learning in computational neuroscience (called error neurons~\cite{PC-1999,PCsamp-2024,PC-2024}). This speed-adjusted synaptic plasticity can be tested in a biological context, e.g., changes in postsynaptic firing rate induces synaptic plasticity observed in previous works~\cite{Xie-1999, Senn-2014, Senn-2018}, \blue{especially the dendritic predictive plasticity reducing the prediction error of somatic spiking in the recent works~\cite{Senn-2014,Senn-2018}}. \blue{We remark that to estimate $h_k$, we only need to collect the contributions from the neighbors of neuron $k$ (this locality is more evident when a sparse network is considered, while our dense net setting can be easily generalized to the sparse network, left as a future work).}

Learning proceeds as a gradient descent:
\begin{equation}\label{eq:update_rule}
%\begin{aligned}
    (J_{kl})_{n+1} = (J_{kl})_n - \eta \frac{\partial \mathcal{L}}{\partial J_{kl}},
%\end{aligned}
\end{equation}
where $\eta > 0$ is the learning rate, and $(J_{kl})_n$ denotes the value of the parameter at iteration $n$. This update rule adjusts $J_{kl}$ to reduce the discrepancy between statistics computed from the model-generated samples (\textit{the model phase}) and those computed from the training data (\textit{the data phase}). The model phase can be estimated by running the Langevin dynamics and collecting intermediate states as samples based on the current estimates of $\{J_{kl}\}$.
The update rule for bias can be derived in a similar way (see the Appendix~\ref{SM-a}). We conclude that NLM turns the classic random recurrent neural network (no learning) into a predictive-learning-driven Langevin dynamics.

\section{Training protocol}\label{train}
In the data phase,
the expectation $\left\langle (x_k - h_k) \phi(x_l) \right\rangle_{data}$ is estimated empirically using a mini-batch of $m$ samples $\{\mathbf{x}^{(s)}\}_{s=1}^{m}$ drawn from the training dataset $\mathbf{X}_{data}$.
Pixels correspond to firing rates in this paper. For each sample $\mathbf{x}^{(s)}$ in the batch, we compute $h_k^{(s)} = \sum_l J_{kl} \phi(x_l^{(s)})+b_k$. The data-dependent term $A_{kl}$ is estimated as
\begin{equation}
    A_{kl} = \beta \cdot \frac{1}{m} \sum_{s=1}^{m} \left( x_k^{(s)} -h_k^{(s)} \right) \tanh(x_l^{(s)}),
\end{equation}
where the notation $x_k$ denotes the $k$-th component of vector $\mathbf{x}$.

In the model phase,
the expectation $\left\langle (x_k - h_k) \phi(x_l) \right\rangle_{model}$ is achieved by simulating the Langevin dynamics described in Eq.~\eqref{eq:langevin}. For practical implementation, the stochastic differential equation is discretized using an Euler-Maruyama scheme with a time step size $dt$:
\begin{equation}
    \mathbf{x}_{t+dt} = \mathbf{x}_t + dt \cdot {\bf{F}}(\mathbf{x}_t) + \sqrt{2T dt} \boldsymbol{\epsilon}_t,
    \label{eq:langevin_discrete}
\end{equation}
where $\mathbf{F}(\mathbf{x}_t) = -\nabla_{\mathbf{x}} E(\mathbf{x}_t)$ is the deterministic force vector with components $F_i$ given by Eq.~\eqref{eq:force_comp}, and $\boldsymbol{\epsilon}_t$ is a vector of independent standard Gaussian random variables.

To improve the sampling efficiency and quality,  $M$ persistent chains (model states $\matr{X}_{model}$) are created. These chains are initialized once (e.g., from a Gaussian distribution) and are then updated by running the Langevin dynamics for $k$ steps. The final states of these chains after $k$ steps are used to estimate the model expectation. The term $B_{kl}$ represents this expectation, scaled by $\beta$:
\begin{equation}
    B_{kl} = \beta \cdot \frac{1}{M} \sum_{s=1}^{M} \left( x_k^{(s)} -h_k^{(s)} \right) \tanh(x_l^{(s)}),
\end{equation}
where $\{\mathbf{x}^{(s)}\}_{s=1}^{M}$ are the samples from all the persistent chains. This term approximates $\beta \left\langle (x_k - h_k) \phi(x_l) \right\rangle_{model}$. After each gradient calculation (learning), the final states of the chains serve as initial states for the next iteration. The process continues until a total number of parameter updates $t_{age}$ is reached. This is the PCD-$k$ algorithm~\cite{PCD-2008} (see the next paragraph). \blue{A single persistent chain with exponential averaging is also possible (see further discussions in Appendix~\ref{SM-d})}. All training protocols are sketched in the Appendix together with their hyperparameters.

To study the out-of-equilibrium regime of NLM training, we consider the following training protocols. First, persistent contrastive divergence (PCD) is used in the learning phase to collect
samples from Langevin dynamics to estimate the model-dependent terms for gradients. $M$ persistent chains are considered, each of them starting from an
independent random initialization [e.g., $\mathbf{x}\sim\mathcal{N}(0,\mathbf{I}_N)$]. For each chain, every $k$ step a sample is collected (i.e., PCD-$k$). We denote $t_{\rm age}$
as the total number of parameter updates. During the generation phase, $t_G$ steps of the Langevin dynamics from a random initialization are iterated before an image is
obtained.  Second, the Rdm-$k$ approach is used as in Ref.~\cite{Decelle-2022}. Different from PCD, at every gradient estimation step, each Langevin chain is re-initialized from a purely random configuration (standard Gaussian noise). The parameter $k$ indicates the exact number of steps performed before a sample is collected for the model-dependent term.
Third, we first train an autoencoder (AE) to compress the images into a low-dimensional latent space where NLM is next trained using these compressed latent vectors. We call this protocol AE-NLM, where the PCD-10 strategy is used in the latent training. In spirit, the AE-NLM is similar to the latent diffusion model, where the diffusion model is trained in the latent space of a pretrained AE~\cite{LDM-2022}.  After the NLM is well-trained in the latent space, the Langevin dynamics starting from a random initialization are run up to $t_G$ steps for the latent sample generation, which is then decoded into a real image using the decoder part of the pretrained AE network. AE-NLM makes the learning occur only in the low-dimensional latent space and thus significantly reduces the computational complexity and enhances the generation quality. All technical details of these three protocols are given in the Appendix~\ref{SM-b}.

\section{Generation process}\label{gen}
The generation process aims to draw samples from the learned model distribution $P(\mathbf{x})$.
This is achieved by simulating the neural Langevin dynamics [Eq. \eqref{eq:langevin}], using the final trained parameters $\matr{J}_{\text{final}}$ and the same temperature:
\begin{equation}
    \frac{d \mathbf{x}}{d t} = -\nabla_{\mathbf{x}} E(\mathbf{x} | \matr{J}_{\text{final}}) + \sqrt{2T} \boldsymbol{\epsilon}(t), \label{eq:langevin_generation}
\end{equation}
where $E(\mathbf{x} | \matr{J}_{\text{final}})$ explicitly denotes the energy function parameterized by the trained $\matr{J}_{\text{final}}$.  Samples $\mathbf{x}$ are guaranteed to be distributed according to $P(\mathbf{x}) = \frac{1}{Z} \exp(-\beta E(\mathbf{x} | \matr{J}_{\text{final}}))$.

The simulation starts from a batch of initial states sampled from $\mathcal{N}(0,\mathbf{I}_N)$ and is run for a specified number of steps, $t_G$, whose detailed procedure is outlined in the Appendix~\ref{SM-b}. Interestingly, the choice of $(k, t_{age}, t_G)$ affects the generative performance, yielding an out-of-equilibrium (OOE) regime best for the performance~\cite{YN-2019,Decelle-2022}. To fully explore these effects, we also consider other training protocols, such as Rdm-$k$ and AE-NLM (see details in the Appendix). NLM also displays memorization-to-generalization transition as training data size increases, evaluated by adversarial accuracy indicators (see the next section). We will discuss these two properties below.

 \begin{figure}
\centering
\includegraphics[width=0.5\textwidth]{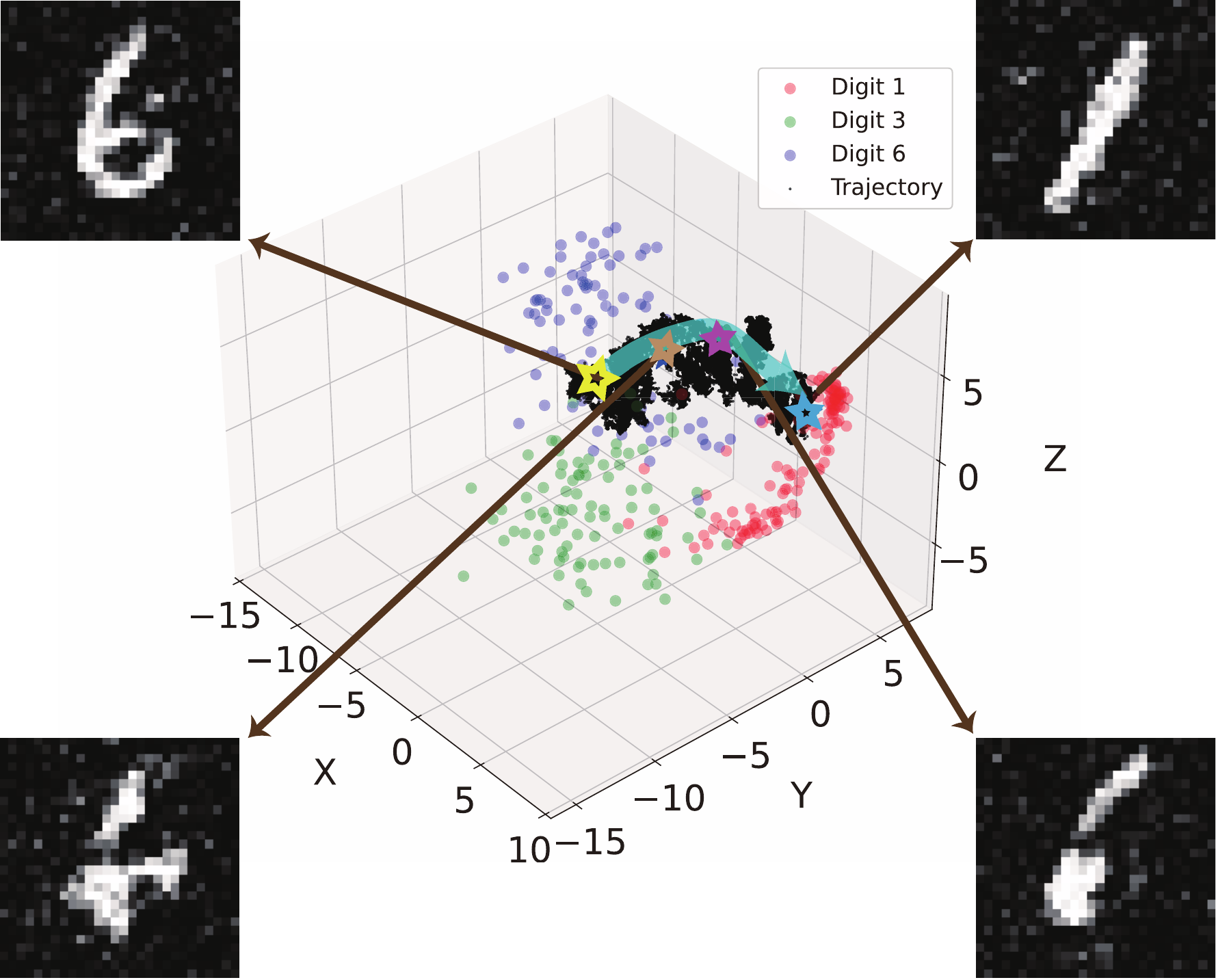}
\caption{State transition during image generation ($T=1$). 
$288$ data points from the training set are considered, which consists of digits 1, 3, and 6, projected onto three-dimensional space via Principal Component Analysis. These principal components were then used as a basis to plot the trajectory of the sampling process. The visualization demonstrates transitions among generated images from digit 6 to digit 1. In the training set, digits 1, 3, and 6 are indicated by red, green, and blue colors, respectively. The sampling trajectory, running from the $120\,000$th step to the $150\,000$th step, is marked by small black dots. PCD-10 is used for training.}
\label{fig1}
\end{figure}

\begin{figure*}
\centering
\includegraphics[width=0.9\textwidth]{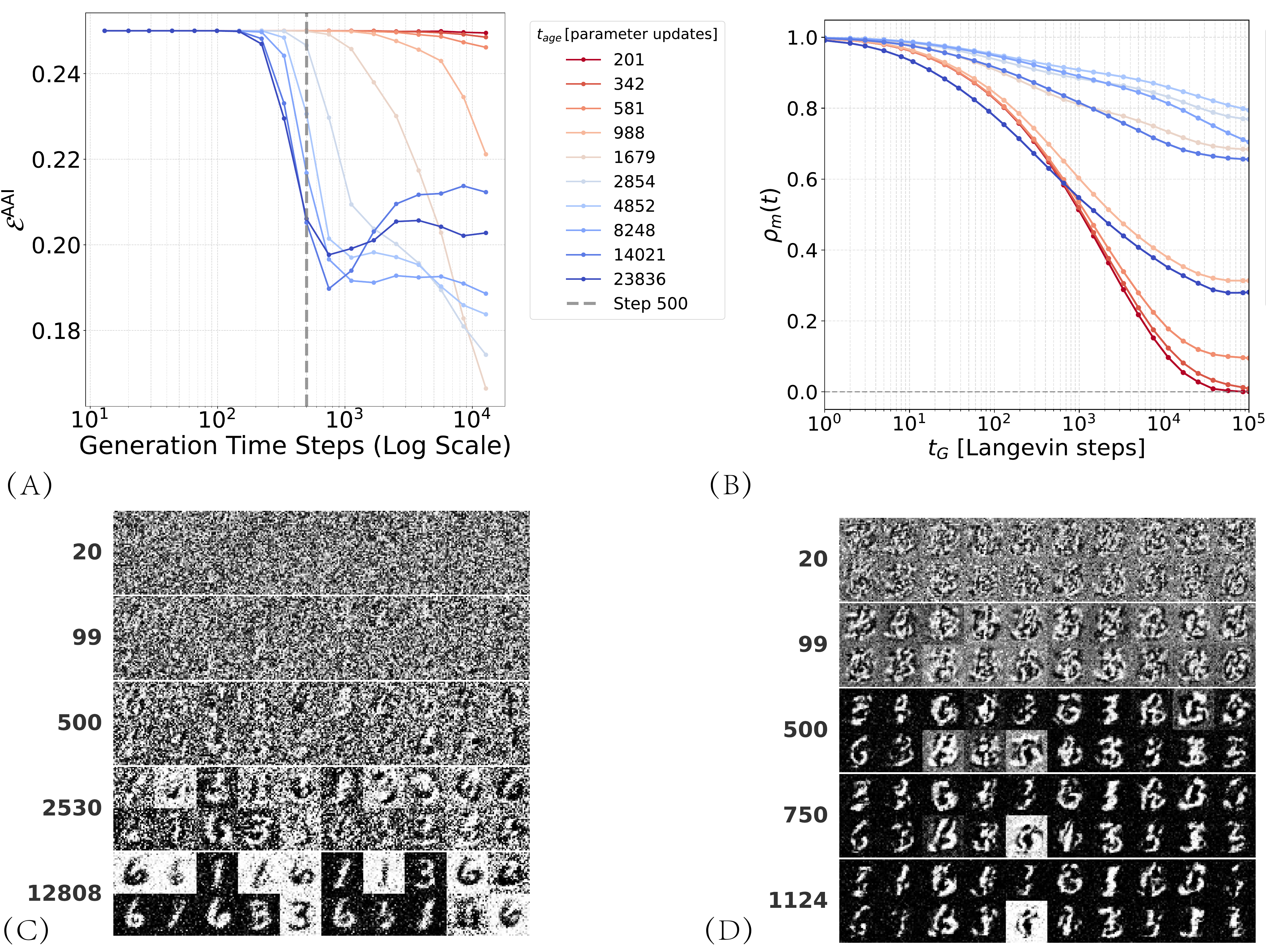}
\caption{Out-of-equilibrium learning regime of Rdm-500. (A) Generative performance measured by ${\cal E}^{{\rm{AAI}}}$ of Rdm-500. (B) Auto-correlation function of magnetization as a function of time $t$, from which the mixing or relaxation time scale $t_{\rm R}$ can be extracted as $\rho_m(t)\sim e^{-t/t_{\rm R}}$. (C) Examples of generated images at different $t_G$ using the learned parameters from $t_{\rm age}=1679$. (D) The same as (C) but $t_{\rm age}=14021$. }
\label{fig2}
\end{figure*}

\section{Evaluation metrics}
To evaluate the generation quality of NLM and detect memorization-to-generalization transition, we adopt the adversarial accuracy indicator (AAI) used in previous works~\cite{Decelle-2022}. AAI measures how well training data and generated data are mixed with each other. To compute AAI, two sets are prepared. One is the target set containing 
only generated samples $\{X_{\rm NLM}^{m}\}_{m=1}^{N_s}$; the other is the source set containing only samples from training data $\{X_{\mathcal{D}}^{m}\}_{m=1}^{N_s}$.
Then we can estimate the probability $Acc_T$ that a generated sample has a nearest neighbor (n.n., \blue{characterized by minimum Euclidean distance but excluding the query sample itself}) which is also a generated sample, and the probability $Acc_S$ that a data sample has a n.n. which is also a data sample. When $N_s$ is large and the model is well-trained, both $Acc_S$ and $Acc_T$ converge to $0.5$, indicating that the generated samples are statistically indistinguishable from
real ones. Therefore, ${\cal E}^{\rm AAI}=0.5[(Acc_S-0.5)^2+(Acc_T-0.5)^2]$. If $Acc_S$ or $Acc_T$ gets close to zero, a memorization is identified. If $Acc_T$ gets close to one, the generated samples are very close to each other, indicating a lack of diversity or well-separation from the real data. More details are presented in the Appendix~\ref{SM-c}.

 \section{Results}
As a proof of concept, we applied PCD-10 to the benchmark dataset of MNIST handwritten digits~\cite{mnist}. For simplicity, we focus on the training set containing three classes of digits in most situations, unless otherwise stated. 
We plot a low-dimensional visualization
of training and generated digits in Fig.~\ref{fig1}, which shows how the generated images change as the neural dynamics evolve on a low-kinetic-energy landscape. This reveals 
the detailed trajectory-level description of the generation process. Evidence was found in the brain about how low-dimensional attractors are created during perception, memory, and decision making~\cite{Nature-2022}. The generative state transition is driven here by the stochastic noise, where every neuron exhibits fluctuations, but all neurons self-organized to a coherent digit. In this example, we consider only three types of digits. The generation quality deteriorates as the number of types increases (see the Appendix~\ref{SM-d}), whereas this can be mitigated by applying AE-NLM (see the following Fashion MNIST example or MNIST in the Appendix~\ref{SM-d}).

\begin{figure*}
    \centering
    \includegraphics[width=0.9\textwidth]{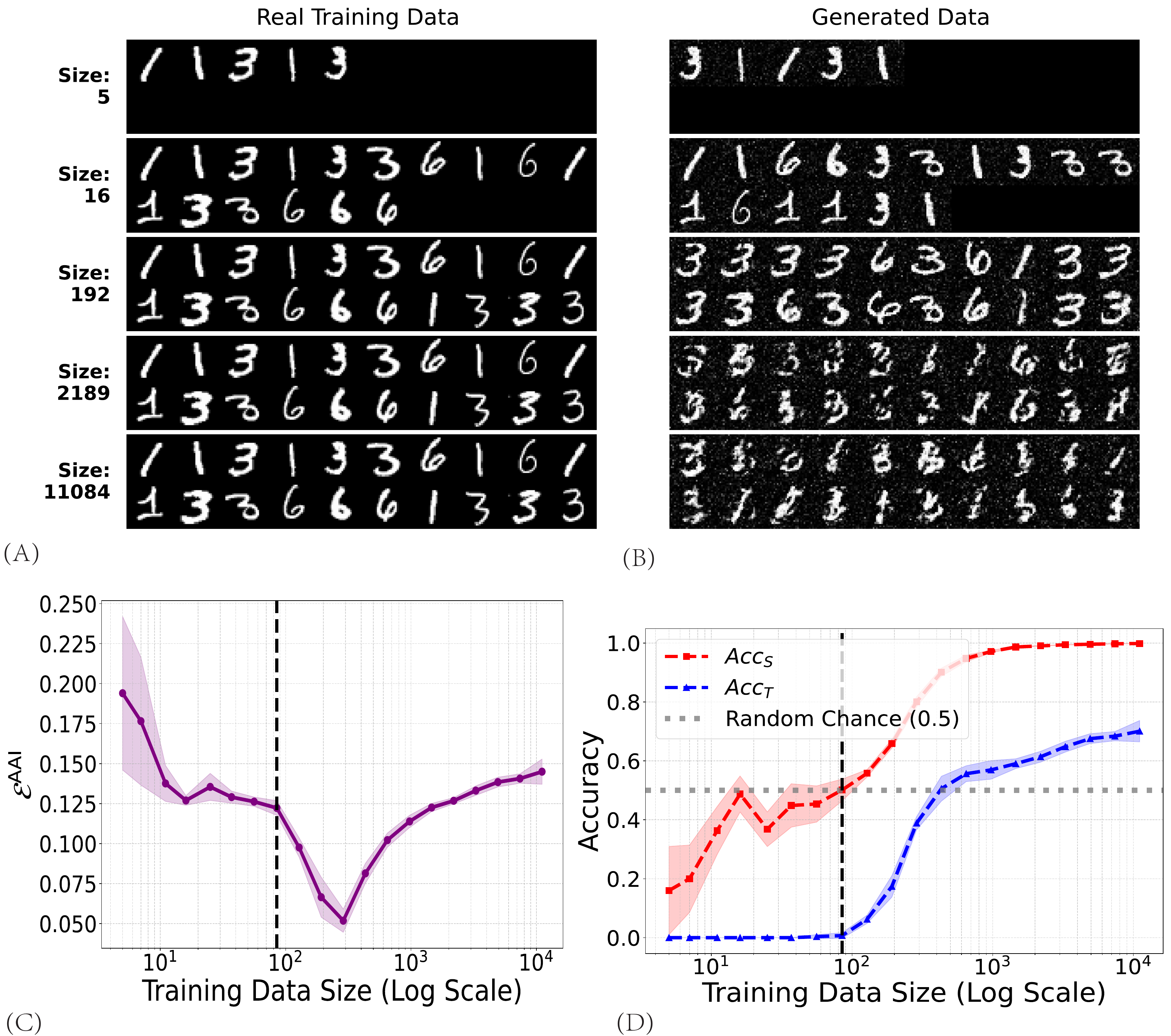}
    \caption{Phase transition from memorization to generalization (PCD-10 used, $t_{\rm age}=20000$, and $t_G=50000$).
    \textbf{(A)} Samples from the training data of different sizes. 
    \textbf{(B)} The corresponding generated samples obtained by NLM trained on the datasets in (A). In the data-deficient regime (e.g., size 5 and 16), the generated images in (B) are pure memorization. 
    \textbf{(C)} ${\cal E}^{{\rm{AAI}}}$ against different data sizes. 
     \textbf{(D)} Separate contribution $Acc_S$ and $Acc_T$ against training data size. The dashed line locates the memorization-to-generalization transition.}
    \label{fig3}
\end{figure*}

\begin{figure}
    \centering
    \includegraphics[width=0.95\textwidth]{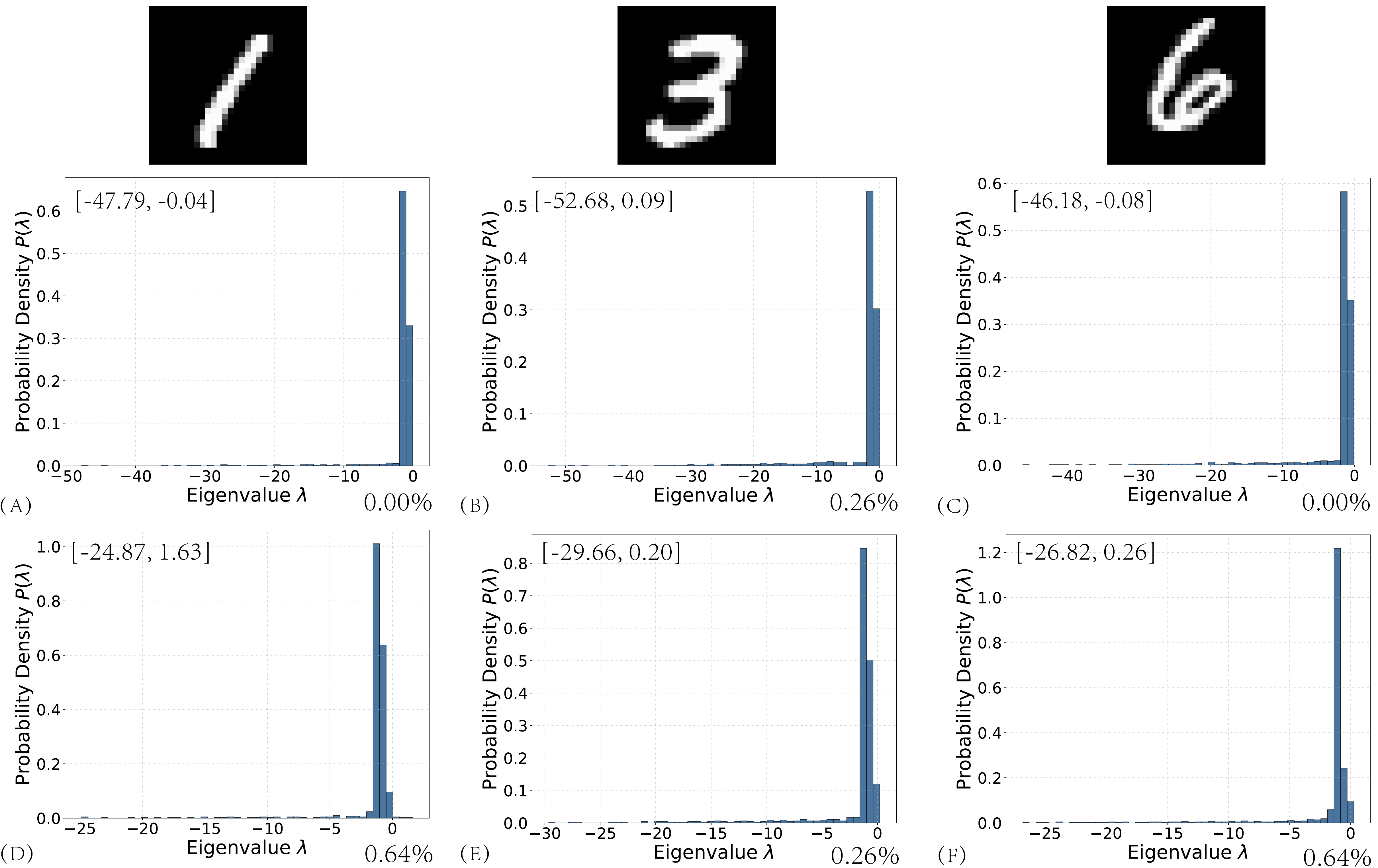}
    \caption{The dynamics Jacobian spectrum (real-valued eigenvalues) for the digit attractors (digits $1$, $6$ and $3$ from training set, shown on the top of the figure) for two training data sizes $16$ (A, B and C) and $192$ (D, E and F). The range of eigenvalues together with the fraction of positive eigenvalues are shown on the plot (top left and bottom right corners, respectively). These two data sizes are also used in Fig.~\ref{fig3}. }
    \label{fig4}
\end{figure}

\begin{figure}
\centering
\includegraphics[width=0.9\textwidth]{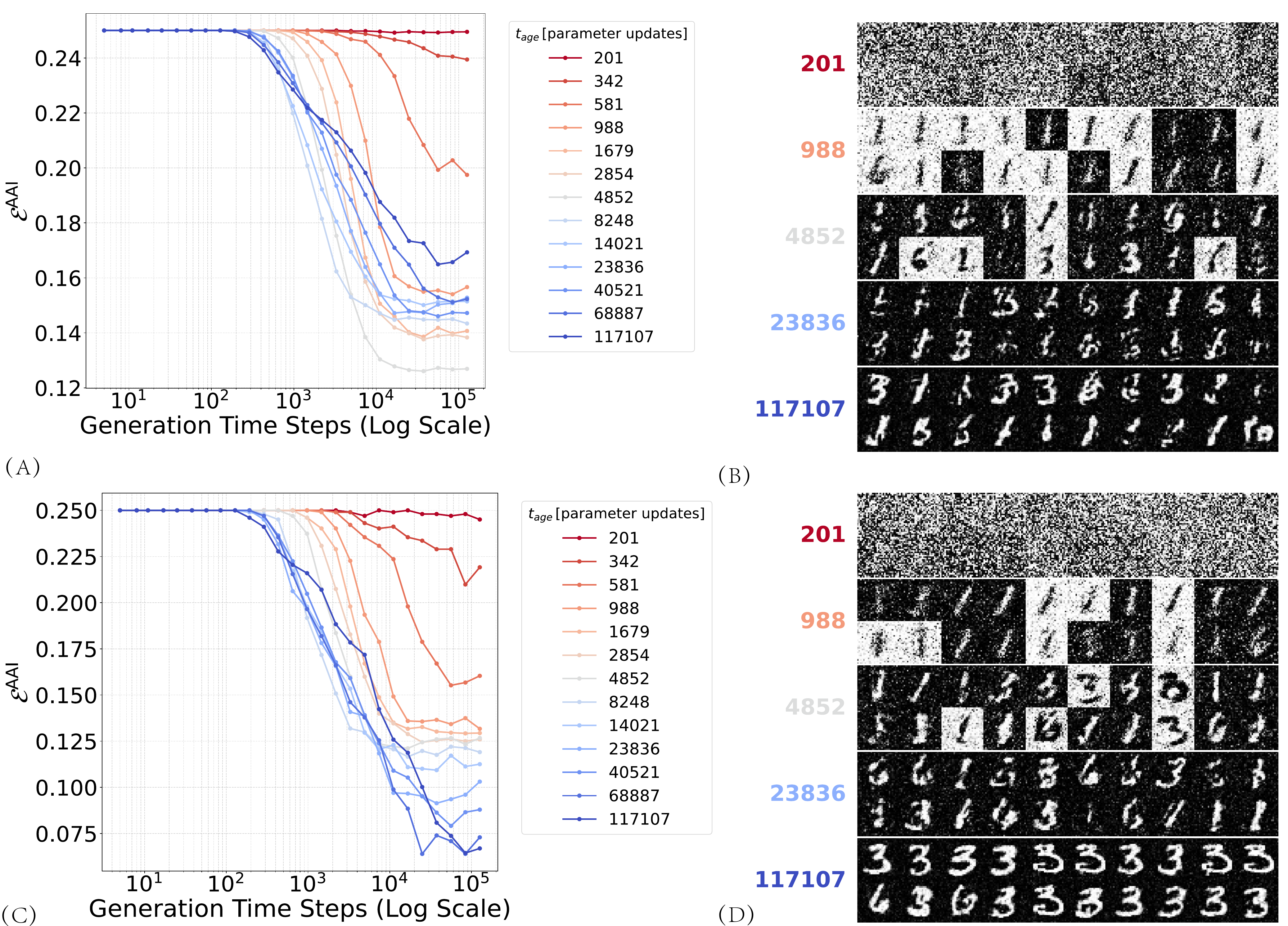}
\caption{Effects of $t_{\rm age}$ on the generative performance of different training data sizes. The generative performance is evaluated by AAI (see the main text and appendix~\ref{SM-c}), and examples of generated digits are also displayed. (A) and (B) for data size $10000$; (C) and (D) for data size $500$. PCD-10 is used for training. }
\label{fig5}
\end{figure}

We then study the OOE effects of NLM, using the Rdm-500 training protocol. Around $t_G=500$, the best performance is achieved (Fig.~\ref{fig2}) \blue{for older machines (trained for a large $t_{\rm age}$)}. Because the older machine (\blue{e.g., its training time $t_{age}$ is larger, but smaller than $8248$}), the longer mixing time~\cite{Decelle-2022} (see also Fig.~\ref{fig2}), the OOE regime can generate good samples without waiting for equilibrium for both training and generation. \blue{In contrast, a large value of $t_G$ for the trained model under Rdm-500 leads to generated samples of less diversity (so-called mode collapse)}. \blue{For a younger machine, a large $t_G$ is able to reach the equilibrium regime (the relaxation time is relatively short), where the quality of generation is also guaranteed (Fig.~\ref{fig2}). The relaxation or mixing time is measured by computing the scaled auto-correlation function: $\rho_m(t)=\frac{C_m(t)}{C_m(0)}$,
where $C_m(t)=\frac{1}{N}\sum_i(\langle\phi_i(x_i(t))\rangle-\bar{m}_i)(\langle\phi_i(x_i(0))\rangle-\bar{m}_i)$ in which $\langle\cdot\rangle$ means the average over multiple Langevin chains [the number is set to $10^4$ in Fig.~\ref{fig2} (B)] after a transient period (e.g., $10^4$ steps here), and $\bar{m}_i\equiv\langle\phi_i(x_i(t_\ell))\rangle_{\ell}$ in which $t_\ell$ denotes the last step of $\ell$-th long ($10^5$ steps here) chain.}

\begin{figure}
\centering
\includegraphics[width=0.9\textwidth]{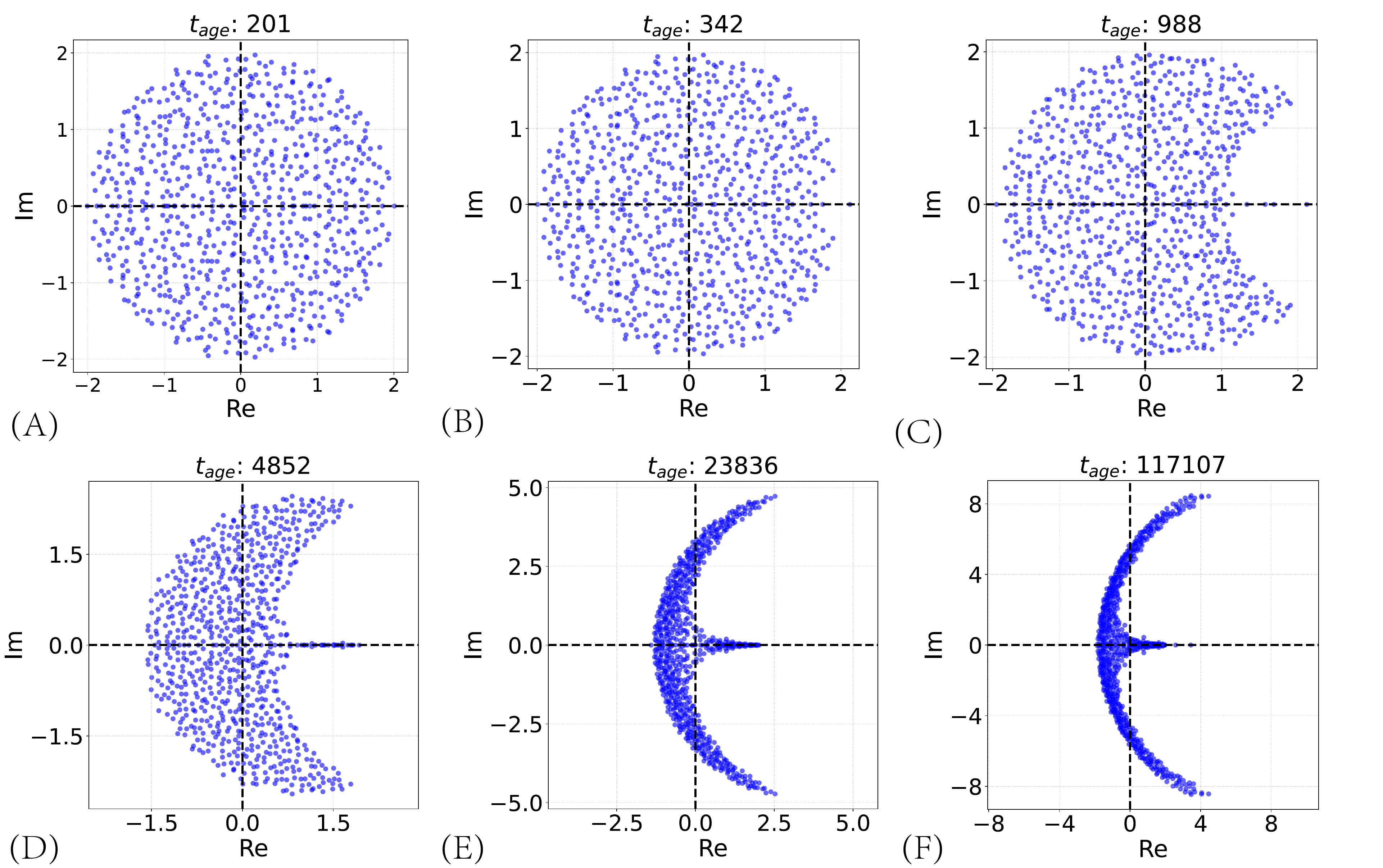}
\caption{Eigenvalue spectrum of $\mathbf{J}$ obtained by PCD-10. $N_{\rm data}=10000$. Different training stages ($t_{\rm age}$) are shown. }
\label{fig6}
\end{figure}

\begin{figure}
\centering
\includegraphics[width=0.9\textwidth]{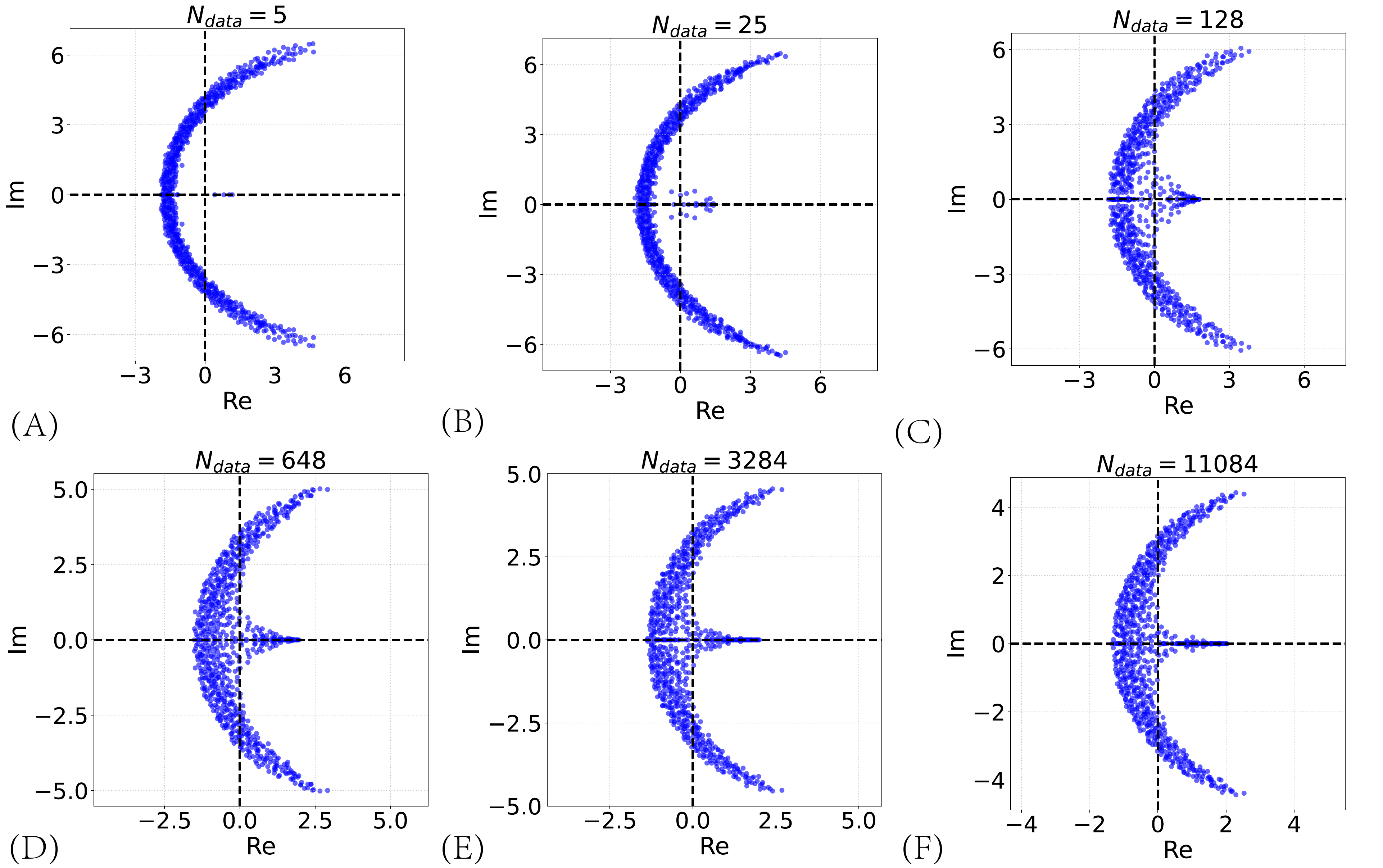}
\caption{Eigenvalue spectrum of $\mathbf{J}$ obtained by PCD-10. Different training data sizes are considered ($t_{\rm age}=20000$). }
\label{fig7}
\end{figure}

We next reveal memorization-to-generalization in NLM. When the training data size is small (smaller than the value indicated by the dashed line), $Acc_T$ vanishes, indicating that a strong overfitting (memorization) occurs, which is also evident when comparing panels (B) and (A) in Fig.~\ref{fig3}. As the data size crosses the threshold (less than $10^2$), NLM becomes creative, generating good samples statistically indistinguishable from real ones. But further increasing the data size will sacrifice the quality, as $Acc_S$ or $Acc_T$ keeps growing until saturation. Other values of $t_{\rm age}$ and $t_G$ do not change the qualitative picture. \blue{By inspecting the dynamics Jacobian spectrum (the Jacobian matrix is defined by $\frac{\partial F_i}{\partial x_j}(\mathbf{x}^*)$, where $\mathbf{x}^*$ is a point near to a training image), we found that the memorization attractors can be destablized towards the generalization attractors (see Fig.~\ref{fig4}). We also observe that in the generalization regime, as the training data size grows, the generative performance gets worse. This is because, for different data sizes, the optimal $t_{\rm age}$ becomes different, as shown in Fig.~\ref{fig5}, while $t_{\rm age}$ and $t_G$ are fixed in Fig.~\ref{fig3} for all data sizes.}

\begin{figure}
    \centering
    \includegraphics[width=0.9\textwidth]{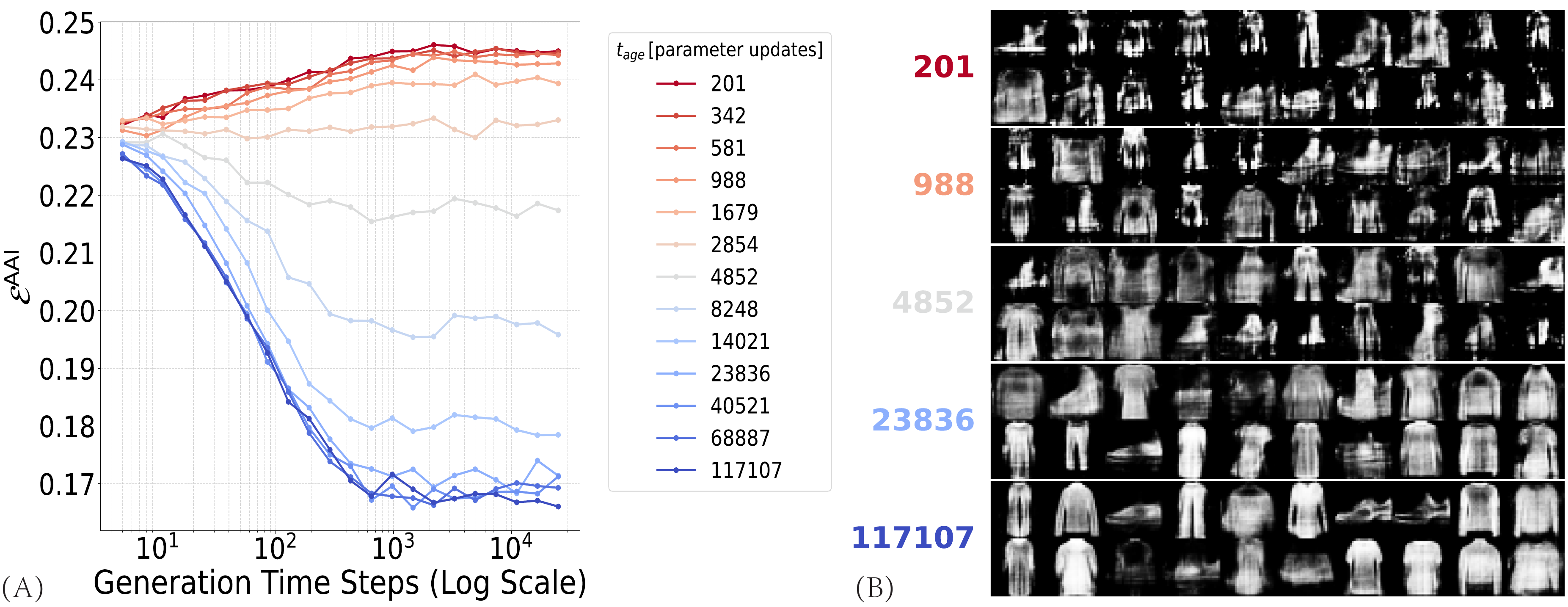} 
    \caption{Generative performance of AE-NLM on the Fashion MNIST dataset (ten categories)~\cite{Fashion-2017} as the training age ($t_{\rm age}$) varies. 
    \textbf{(A)} AAI as a function of $t_{G}$. 
    \textbf{(B)} Generated samples derived from the $30$-dimensional latent space, where recurrent dynamics are sampled. Different training ages are considered, and $t_{G}=24939$.
    }
    \label{fig8}
\end{figure}

We also find that during learning the spectrum of training $\mathbf{J}$ is initially isotropic, and then stretched along both real and imaginary axes, and finally shaped as an anchor (Fig.~\ref{fig6}), while increasing data size at fixed $t_{\rm age}$, the shape of eigenvalue spectrum changes from a sickle to an anchor (Fig.~\ref{fig7}), suggesting that eigenvalues distributed around the positive real axis play a key role for the generalization behavior.

We finally remark that the current framework is an energy-based recurrent dynamics, with fixed points reachable, fitting the algorithmic setting of equilibrium propagation~\cite{EP-2017}, thereby opening up the possibility of latent space reasoning. Moreover, combined with encoder and decoder structures~\cite{Diffvae-2022,LDM-2022}, the latent space neural dynamics can enhance the quality of generation (as shown in Fig.~\ref{fig8}), which deserves further studies. \blue{Note that when $t_G$ is small, the performance is bad, demonstrating that the recurrent latent dynamics via NLM forms generative attractors and before reaching the attractors, the generative performance is not guaranteed even if the encoder and decoder are both well-trained in the auto-encoder structure. The latent NLM thus plays a key role in the performance of AE-NLM, demonstrating the latent space training deserves further studies for relating synaptic learning to neural capability.}

\begin{figure}
\centering
\includegraphics[width=0.9\textwidth]{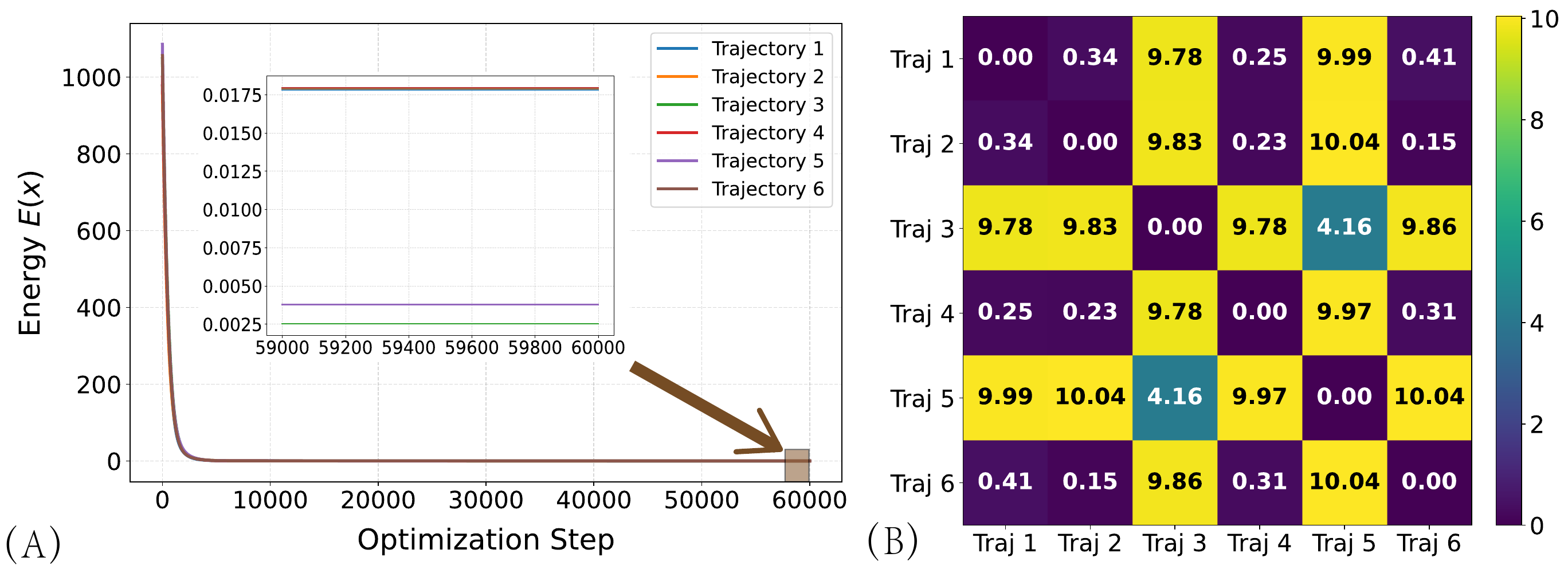}
\caption{Energy evolution of sample trajectories during the fixed-point search. Six typical trajectories of a learned NLM are shown in (A).
Fixed point is detected once the velocity norm $\Vert\dot{\mathbf{x}}\Vert_2<0.001$. When considering $1000$ initializations for the Langevin dynamics, we observed $901$ distinct fixed points with pairwise Euclidean distance larger than $0.01$. A Euclidean-distance matrix is shown in (B).
The actual number of fixed points should be larger than $901$. The training details: $N_{\rm data}=10000$, $t_{\rm age}=4852$, and PCD-10 is used. }
\label{fig9}
\end{figure}

 It is well-known that recurrent neural networks with asymmetric couplings produce an \textit{exponential} number of unstable fixed points~\cite{Helias-2025}. \blue{Although synaptic couplings in a learned NLM are usually non-random and not Gaussian distributed, our framework is able to learn the set of generative fixed points. In fact, we empirically study the set of fixed points in a learned NLM, and we found a large number of fixed points even considering $1000$ random initializations for the Langevin dynamics (Fig.~\ref{fig9}). Although we could not find all fixed points for this large practical network, it is reasonable to conclude from this empirical observation that many fixed points exist in a learned NLM.} Our work further shows that these fixed points can also be used for denoising corrupted images, which is a new sort of associative memory with asymmetric coupling (Fig.~\ref{fig10}, training details are presented in appendix~\ref{SM-e}). We show a proof of concept here, and leave a systematic study of this promising direction to ongoing works. 

\begin{figure}
\centering
\includegraphics[width=0.6\textwidth]{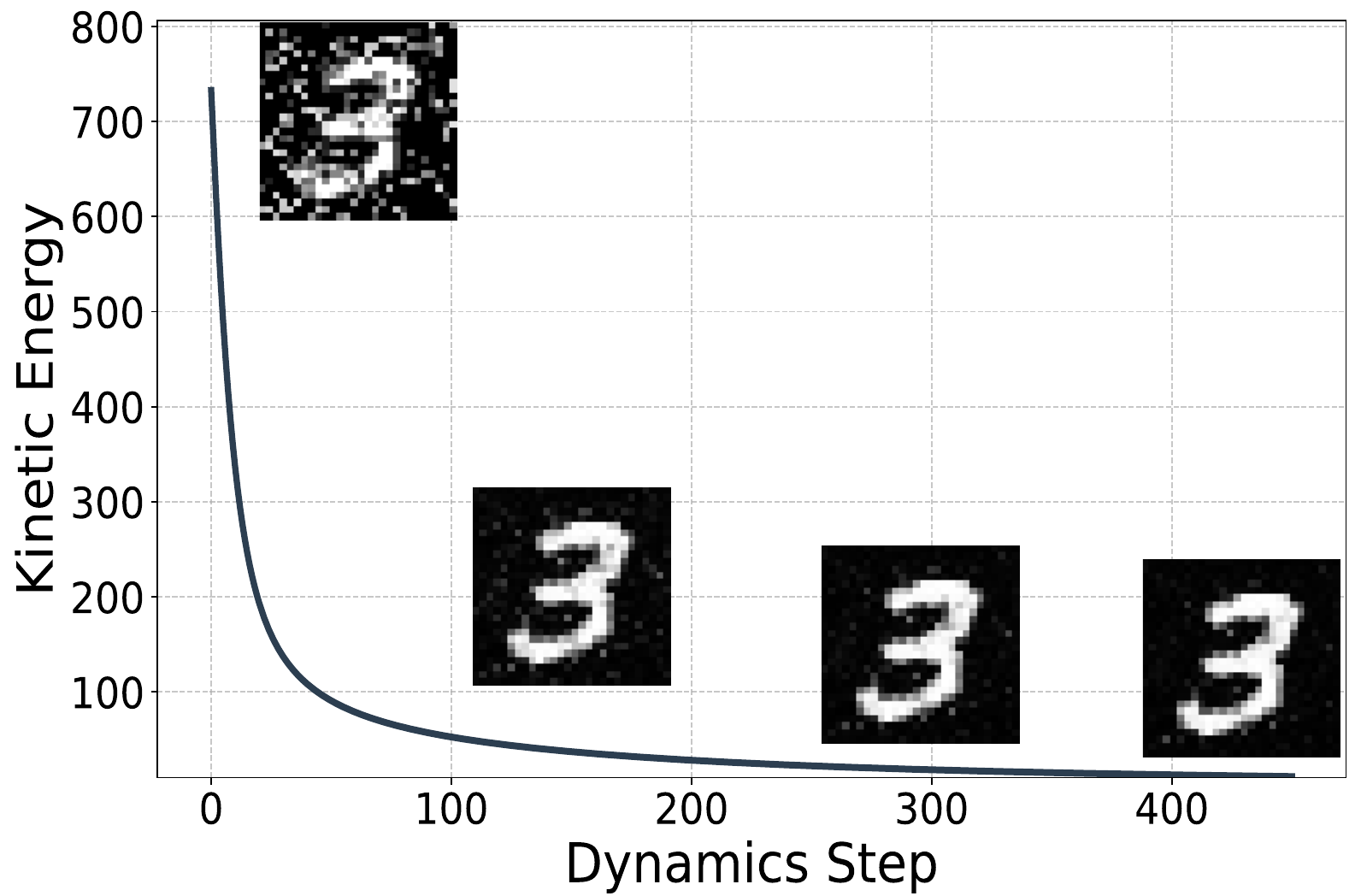}
\caption{ NLM as an associative memory network. Network parameters are trained to store the given dataset of 10 digits. After training, $25\%$ of pixels are replaced by a uniform random variable from $[-1,1]$. NLM starting from this corrupted state
is iterated to retrieval the clean digit (3). Training details are given in the appendix~\ref{SM-e}. }
\label{fig10}
\end{figure}

\section{Conclusion and future remarks}
In this work, we propose a neural Langevin machine that reveals a hidden computational merit of previously discovered unstable fixed points in random recurrent neural networks. Learning makes these unstable fixed points reshaped but reachable and thus usable for memory and generalization. For this generative model, we derive a local learning rule for asymmetric couplings (beyond the simple symmetric Hebbian plasticity rule~\cite{Plast-2020}) and realize state transition among novel generated images, revealing the OOE regime of data generation and further memorization-to-generalization transition as the training data size grows. Future directions include an improvement of data quality in generated samples through latent space training, latent space reasoning using these biologically plausible recurrent dynamics (\blue{our gradient dynamics with asymmetric couplings can be expressed into the form of apical-basal integration~\cite{Senn-2014,LE-2021,Senn-2024}}), and how the NLM learning rule is related to predictive learning, especially when applied to hierarchically coupled systems and excitatory-inhibitory sparse biological networks \blue{with dendritic structures}. We thus anticipate that this NLM will inspire further developments in the field of generative models, thereby guiding us to understand how we subjectively perceive, memorize, and imagine the objective outside world.

%%%%%%%%%%%%%%%%%%%%%%%%%%%%%%%%%%%%%%%%%%%%%%%%%%%%%%%
\section*{Acknowledgments}
This research was supported by the National Natural Science Foundation of China for
Grant number 12475045, and Guangdong Provincial Key Laboratory of Magnetoelectric Physics and Devices (No. 2022B1212010008), and Guangdong Basic and Applied Basic Research Foundation (Grant No. 2023B1515040023).
%%%%%%%%%%%%%%%%%%%%%%%%%%%%%%%%%%%%%%%%%%%%%%%%%%%%%%%%%%%%%%%
\section*{Data availability}
Codes are available in our GitHub~\cite{code-2025}.
\section*{Use of generative-AI tools}
Generative-AI tools such as Gemini 3.8 Flash, DeepSeek-R1, Kimi K3 are used for preparing codes in this paper. The authors examined and verified the output of the AI-generated codes. The authors take full responsibility for the scientific content of this work.
%\newpage
%\onecolumngrid
%\begin{widetext}
\appendix

\section{Detailed derivation of learning rules}\label{SM-a}
In this section, we derive the learning rules under the minimal entropy principle in detail.
To perform gradient-based optimization, we compute the partial derivative of $\mathcal{L}_{KL}$ [or effectively $\mathcal{L}(\J)$] with respect to a specific parameter $J_{kl}$:
\begin{equation}
\begin{aligned}
    \frac{\partial \mathcal{L}_{KL}}{\partial J_{kl}} &= \frac{\partial}{\partial J_{kl}} \left( \ln Z + \beta \langle E(\mathbf{x}) \rangle_{data} \right) \\
    &= \frac{1}{Z} \frac{\partial Z}{\partial J_{kl}} + \beta \left\langle \frac{\partial E(\mathbf{x})}{\partial J_{kl}} \right\rangle_{data}. \label{eq:kl_deriv_step1}
\end{aligned}
\end{equation}
The derivative of the partition function is given by
\begin{equation}
\begin{aligned}
    \frac{\partial Z}{\partial J_{kl}} &= \frac{\partial}{\partial J_{kl}} \int \exp(-\beta E(\mathbf{x})) d\mathbf{x} \\
    &= \int -\beta \frac{\partial E(\mathbf{x})}{\partial J_{kl}} \exp(-\beta E(\mathbf{x})) d\mathbf{x}.
\end{aligned}
\end{equation}
Thus, the first term in Eq. \eqref{eq:kl_deriv_step1} becomes:
\begin{equation}
\begin{aligned}
    \frac{1}{Z} \frac{\partial Z}{\partial J_{kl}} &= -\beta \int \frac{\partial E(\mathbf{x})}{\partial J_{kl}} \frac{\exp(-\beta E(\mathbf{x}))}{Z} d\mathbf{x}\\
    & = -\beta \left\langle \frac{\partial E(\mathbf{x})}{\partial J_{kl}} \right\rangle_{model},
    \end{aligned}
\end{equation}
where $\langle \cdot \rangle_{model}$ denotes an expectation taken with respect to the model distribution $P(\mathbf{x})$.
Substituting this back into Eq. \eqref{eq:kl_deriv_step1}, we get:
\begin{equation}
    \frac{\partial \mathcal{L}_{KL}}{\partial J_{kl}} = -\beta \left\langle \frac{\partial E(\mathbf{x})}{\partial J_{kl}} \right\rangle_{model} + \beta \left\langle \frac{\partial E(\mathbf{x})}{\partial J_{kl}} \right\rangle_{data}. \label{eq:kl_deriv_intermediate}
\end{equation}
Equation~\eqref{eq:kl_deriv_intermediate} implies that when the two gradients of the Hamiltonian match with each other, the learning terminates.

Next, we need the derivative of the energy function $E(\mathbf{x})$ [see Eq. (2) in the main text] with respect to $J_{kl}$:
\begin{equation}\label{eq:energy_deriv_jkl}
    \frac{\partial E(\mathbf{x})}{\partial J_{kl}} = (-x_k + h_k) \phi(x_l), 
\end{equation}
where $h_k = \sum_{m=1}^N J_{km} \phi(x_m)+b_k$ is defined as in the main text. 

Substituting Eq. \eqref{eq:energy_deriv_jkl} into Eq. \eqref{eq:kl_deriv_intermediate}, we obtain:
\begin{equation}\label{eq:kl_grad_final}
%\begin{aligned}
    \frac{\partial \mathcal{L}_{KL}}{\partial J_{kl}} 
    = \beta \left[ \left\langle (x_k - h_k) \phi(x_l) \right\rangle_{model} - \left\langle (x_k - h_k) \phi(x_l) \right\rangle_{data} \right],
%\end{aligned}
\end{equation}
where $\beta = 1/T$. 

Similarly, we compute the derivative of the energy function with respect to the bias $b_k$:
\begin{align}
    \frac{\partial E(\mathbf{x})}{\partial b_k} &= \frac{\partial}{\partial b_k} \left[ \frac{1}{2}\sum_{i=1}^N \left( -x_i + \sum_{m=1}^N J_{im} \phi(x_m) + b_i \right)^2 \right] \nonumber \\
    &= \left( -x_k + \sum_{m=1}^N J_{km} \phi(x_m) + b_k \right) \cdot 1 \nonumber \\
    &= (-x_k + h_k ).
\end{align}
This provides a simple and intuitive learning rule for the bias $b_k$, which aims to match the average value of the term $(-x_k + h_k)$ across data and model samples:
\begin{equation}
    \frac{\partial \mathcal{L}_{KL}}{\partial b_k} = \beta \left[ \left\langle (x_k - h_k) \right\rangle_{model} - \left\langle (x_k - h_k) \right\rangle_{data}\right].
\end{equation}

The primary goal of the training process is to adjust the coupling parameters $\matr{J}$ (or $\mathbf{b}$) of the quasi-potential model such that its equilibrium distribution $P(\mathbf{x})$ closely approximates the target data distribution $P_{data}(\mathbf{x})$. This is achieved by minimizing the KL divergence $\mathcal{L}_{KL}(P_{data} || P)$ with respect to $\matr{J}$ (or $\mathbf{b}$). The optimization is performed using stochastic gradient descent, where the gradient of the KL divergence with respect to a parameter $J_{kl}$ is given by Eq.~\eqref{eq:kl_grad_final}. The evaluation of this gradient involves two expectation terms: one with respect to the model distribution $P(\mathbf{x})$ and the other with respect to the data distribution $P_{data}(\mathbf{x})$. We thus call this type of learning contrastive learning. Two terms are balanced only if the model distribution gets close to the true one.

\section{Algorithmic details of neural Langevin machine}\label{SM-b}
This section details the algorithmic implementation of the generative model, including both training process and generation process. For simplicity, we focus on the training set containing three classes of digits in most situations, unless otherwise stated. 

\subsection{Training process}
 In the matrix form for a set of states $\matr{X}_t$, the dynamics can thus be written as follows (used in Alg.~\ref{alg:pcd_langevin_training}):
\begin{equation}
 \matr{X}_{t+dt} = (1 - dt) \matr{X}_t + dt(\mathbb{B}+ \matr{J} \tanh(\matr{X}_t)) - dt \cdot (\phi'(\matr{X}_t) \odot (\matr{J}^\top (\matr{J} \tanh(\matr{X}_t)+\mathbb{B} - \matr{X}_t))) + \sqrt{2T \, dt} \, \boldsymbol{\mathcal{E}}_t, 
\end{equation}
where $\mathbb{B}\in\mathbb{R}^{N\times M}$ each column of which is formed by $\mathbf{b}$, $\odot$ denotes Hadamard product, $\phi'(x) = 1 - \tanh^2(x)$, and $\boldsymbol{\mathcal{E}}_t$ is a matrix of i.i.d. Gaussian noise.

To improve the sampling efficiency and quality,  $M$ persistent chains (model states $\matr{X}_{model}$) are created. These chains are initialized once (e.g., from a Gaussian distribution) and then updated by running the Langevin dynamics for $k$ steps. The final states of these chains after $k$ steps are used to estimate the model expectation. The term $B_{kl}$ in Alg.~\ref{alg:pcd_langevin_training} represents this expectation, scaled by $\beta$:
\begin{equation}
    B_{kl} = \beta \cdot \frac{1}{M} \sum_{s=1}^{M} \left( x_k^{(s)}-b_k - \left(\matr{J} \tanh(\mathbf{x}^{(s)})\right)_k \right) \tanh(x_l^{(s)}),
\end{equation}
where $\{\mathbf{x}^{(s)}\}_{s=1}^{M}$ are the samples from all the persistent chains. This term approximates $\beta \left\langle (x_k - h_k) \phi(x_l) \right\rangle_{model}$. After each gradient calculation (learning), the final states of the chains serve as initial states for the next iteration (so-called persistent chains).

With the data-dependent term $A_{kl}$ shown in the main text and model-dependent term $B_{kl}$ computed, the gradient of the KL divergence with respect to $J_{kl}$ is approximated by $(B_{kl} - A_{kl})$. The coupling parameters are then updated using gradient descent:
\begin{equation}
    (J_{kl})_{new} = (J_{kl})_{old} - \eta (B_{kl} - A_{kl}), \label{eq:param_update_alg}
\end{equation}
where $\eta$ (see Alg.~\ref{alg:pcd_langevin_training}) is the learning rate. Additionally, a weight decay term with coefficient $\lambda$ is typically included in the optimization step to regularize the model and prevent overfitting. This corresponds to adding $\lambda J_{kl}$ to the above gradient term. The bias terms are similarly updated. The overall training procedure is sketched in Alg.~\ref{alg:pcd_langevin_training}. The process continues until a total number of parameter updates $t_{age}$ is reached. This type of learning protocol is called persistent contrastive divergence (PCD)~\cite{PCD-2008}. 

\begin{algorithm}
\caption{Training Protocol of PCD-$k$ NLM}
\label{alg:pcd_langevin_training}
\begin{algorithmic}[1]

\State \textbf{Hyperparameters:} $N$: network size; $g$: initial coupling strength; $T$: temperature; $\Delta t$: time increment; $n$: number of epochs; $k$: dynamic steps per update; $\eta$: learning rate; $\epsilon=0.03$; $\lambda$: weight decay parameter; $m$: mini-batch size; $M$: number of persistent chains.

\State \textbf{Initialization:}
\State Load MNIST dataset with target digits and preprocess data to unbounded space: ${\bf{X}}_{data} = \text{atanh}\left(\text{clip}(2 \cdot {\bf{X}}_{raw} - 1, -1+\epsilon, 1-\epsilon)\right)$.
\State Initialize $\matr{J} \in \mathbb{R}^{N \times N}$ from $\mathcal{N}(0, g^2/N)$, and strictly set diagonal $J_{ii} = 0$.
\State Initialize bias vector $\mathbf{b} \in \mathbb{R}^N$ as zeros.
\State Initialize persistent model states $\matr{X}_{model} \in \mathbb{R}^{N \times M}$ drawn from $\mathcal{N}(0, \mathbf{I})$.

\State \textbf{Training Loop:}
\For{epoch $= 1$ to $n$}
    \For{each batch ${\bf{X}}_{batch} \in \mathbb{R}^{N \times m}$ from DataLoader(${\bf{X}}_{data}$)}
        \State \textbf{Data Phase Gradients:}

        \State Compute local field: $\matr{H}_{batch} = \matr{J} \tanh({\bf{X}}_{batch}) + \mathbf{b} \mathbf{1}_{m}^\top$
        \State $\nabla_{\matr{J}}^{data} = \frac{1}{T \cdot m} (-\mathbf{X}_{batch} + \matr{H}_{batch}) \tanh({\bf{X}}_{batch})^\top$
        \State $\nabla_{\mathbf{b}}^{data} = \frac{1}{T \cdot m} (-\mathbf{X}_{batch} + \matr{H}_{batch}) \mathbf{1}_{m}$
        
        \State \textbf{Model Phase (Langevin Dynamics):}
        \State ${\bf{X}}_0 \leftarrow \matr{X}_{model}$
        \For{$s = 1$ to $k$}
            \State $\matr{H}_s = \matr{J} \tanh({\bf{X}}_{s-1}) + \mathbf{b} \mathbf{1}_{M}^\top$
            \State $\boldsymbol{\delta J}_s = \matr{J}^\top (\matr{H}_{s-1} - {\bf{X}}_{s-1})$
            \State $\phi'({\bf{X}}_{s-1}) = 1 - \tanh^2({\bf{X}}_{s-1})$ \Comment{Element-wise}
            \State Draw noise $\boldsymbol{\mathcal{E}}_s \sim \mathcal{N}(0, \mathbf{I})$
            \State ${\bf{X}}_{s} = (1 - \Delta t) {\bf{X}}_{s-1} + \Delta t \cdot \matr{H}_{s} - \Delta t \cdot (\phi'({\bf{X}}_{s-1}) \odot \boldsymbol{\delta J}_{s}) + \sqrt{2T \Delta t} \, \boldsymbol{\mathcal{E}}_{s}$
        \EndFor
        \State Update persistent states $\matr{X}_{model} \leftarrow {\bf{X}}_{k}$.
        
        \State \textbf{Model Phase Gradients:}
        \State $\matr{H}_{model} = \matr{J} \tanh(\matr{X}_{model}) + \mathbf{b} \mathbf{1}_{M}^\top$
        \State $\nabla_{\matr{J}}^{model} = \frac{1}{T \cdot M} (-\matr{X}_{model} + \matr{H}_{model}) \tanh(\matr{X}_{model})^\top$
        \State $\nabla_{\mathbf{b}}^{model} = \frac{1}{T \cdot M} (-\matr{X}_{model} + \matr{H}_{model}) \mathbf{1}_{M}$
        
        \State \textbf{Parameter Update (Gradient Descent):}
        \State Set diagonal $\nabla_{J_{ii}}^{data} - \nabla_{J_{ii}}^{model} = 0$
        \State $\matr{J} \leftarrow \matr{J} - \eta (\nabla_{\matr{J}}^{data} - \nabla_{\matr{J}}^{model} + \lambda \matr{J})$
        
        \State $\mathbf{b} \leftarrow \mathbf{b} - \eta (\nabla_{\mathbf{b}}^{data} - \nabla_{\mathbf{b}}^{model})$ \Comment{No weight decay for bias}
    \EndFor 
\State \textbf{Store} the trained parameters $\mathbf{J}_{\text{trained}}$ and $\mathbf{b}_{\text{trained}}$ if necessary.
\EndFor 
\end{algorithmic}
\end{algorithm}

To investigate the non-equilibrium regimes during the learning process, we adopted the Rdm-$k$ methodology~\cite{Decelle-2022}. Unlike PCD, which maintains persistent stochastic dynamics chains across parameter updates to approximate the model distribution, the Rdm-$k$ approach breaks this temporal dependency. Instead, at every gradient estimation step, the chains are re-initialized from a purely random configuration (standard Gaussian noise). The parameter $k$ dictates the exact number of dynamics steps performed before parameter updates (or gradient estimation). The system thus executes $k$ steps of Langevin dynamics, and solely the configuration at the very end of the chain is utilized to approximate the model expectations $\langle \cdot \rangle_{model}$. The training protocol is detailed in Alg.~\ref{alg:rdmk_langevin_training}.

\begin{algorithm}
\caption{Training Protocol of the Rdm-$k$ NLM}
\label{alg:rdmk_langevin_training}
\begin{algorithmic}[1]

\State \textbf{Hyperparameters:} $N$: network size; $g$: initial coupling strength; $T$: temperature; $\Delta t$: time increment; $n$: number of epochs; $k$:  dynamics steps per update; $\eta$: learning rate; $\lambda$: weight decay parameter; $m$: mini-batch size; $M$: number of parallel chains.

\State \textbf{Initialization:}
\State Load MNIST dataset according to desired types of digits.
\State Preprocess data to unbounded space: ${\bf{X}}_{data} = \text{atanh}\left(\text{clip}(2 \cdot {\bf{X}}_{raw} - 1, -1+\epsilon, 1-\epsilon)\right)$.
\State Initialize $\matr{J} \in \mathbb{R}^{N \times N}$ from $\mathcal{N}(0, g^2/N)$, and strictly set diagonal $J_{ii} = 0$.
\State Initialize bias vector $\mathbf{b} \in \mathbb{R}^N$ as zeros.

\State \textbf{Training Loop:}
\For{epoch $= 1$ to $n$}
    \For{each batch ${\bf{X}}_{batch} \in \mathbb{R}^{N \times m}$ from DataLoader(${\bf{X}}_{data}$)}
        \State \textbf{Data Phase Gradients:}
        \State Compute local field: $\matr{H}_{batch} = \matr{J} \tanh({\bf{X}}_{batch}) + \mathbf{b} \mathbf{1}_{m}^\top$
        \State $\nabla_{\matr{J}}^{data} = \frac{1}{T \cdot m} (-\mathbf{X}_{batch} + \matr{H}_{batch}) \tanh({\bf{X}}_{batch})^\top$
        \State $\nabla_{\mathbf{b}}^{data} = \frac{1}{T \cdot m} (-\mathbf{X}_{batch} + \matr{H}_{batch}) \mathbf{1}_{m}$
        
        \State \textbf{Model Phase (Langevin Dynamics):}
        \State Initialize chains from random noise: ${\bf{X}}_0 \sim \mathcal{N}(0, \mathbf{I})$ \Comment{Core difference from PCD}
        \For{$s = 1$ to $k$}
            \State $\matr{H}_s = \matr{J} \tanh({\bf{X}}_{s-1}) + \mathbf{b} \mathbf{1}_{M}^\top$
            \State $\boldsymbol{\delta J}_s = \matr{J}^\top (\matr{H}_{s-1} - {\bf{X}}_{s-1})$
            \State $\phi'({\bf{X}}_s) = 1 - \tanh^2({\bf{X}}_{s-1})$ \Comment{Element-wise}
            \State Draw noise $\boldsymbol{\mathcal{E}}_s \sim \mathcal{N}(0, \mathbf{I})$
            \State ${\bf{X}}_{s} = (1 - \Delta t) {\bf{X}}_{s-1} + \Delta t \cdot \matr{H}_{s} - \Delta t \cdot (\phi'({\bf{X}}_{s}) \odot \boldsymbol{\delta J}_{s}) + \sqrt{2T \Delta t} \, \boldsymbol{\mathcal{E}}_{s}$
        \EndFor
        \State Set terminal state: $\matr{X}_{model} = {\bf{X}}_{k}$ \Comment{Discard memory for next update}
        
        \State \textbf{Model Phase Gradients:}
        \State $\matr{H}_{model} = \matr{J} \tanh(\matr{X}_{model}) + \mathbf{b} \mathbf{1}_{M}^\top$
        \State $\nabla_{\matr{J}}^{model} = \frac{1}{T \cdot M} (-\matr{X}_{model} + \matr{H}_{model}) \tanh(\matr{X}_{model})^\top$
        \State $\nabla_{\mathbf{b}}^{model} = \frac{1}{T \cdot M} (-\matr{X}_{model} + \matr{H}_{model}) \mathbf{1}_{M}$
        
        \State \textbf{Parameter Update (Gradient Descent):}
        \State Set diagonal $\nabla_{J_{ii}}^{data} - \nabla_{J_{ii}}^{model} = 0$
        \State $\matr{J} \leftarrow \matr{J} - \eta (\nabla_{\matr{J}}^{data} - \nabla_{\matr{J}}^{model} + \lambda \matr{J})$
        \State $\mathbf{b} \leftarrow \mathbf{b} - \eta (\nabla_{\mathbf{b}}^{data} - \nabla_{\mathbf{b}}^{model})$ \Comment{No weight decay for bias}
    \EndFor 
\State \textbf{Store} the trained parameters $\mathbf{J}_{\text{trained}}$ and $\mathbf{b}_{\text{trained}}$ if necessary.
\EndFor 
\end{algorithmic}
\end{algorithm}
 
To enhance the generation quality of NLM, we designed the AE-NLM framework. More precisely, we first independently train an auto-encoder (AE) following standard backpropagation. 
 By encoding the original data into a lower-dimensional manifold, the generative model can focus on learning the fundamental semantic structure of the data while filtering out imperceptible, high-frequency pixel-level noise. This approach substantially improves both training stability and final generation quality. Inspired by this paradigm, we call this strategy AE-NLM. 
 
 The cost function for the independent auto-encoder training is given as a reconstruction loss:
\begin{equation}
    \mathcal{L} = \mathcal{L}_{MSE}(\mathbf{x}, \hat{\mathbf{x}}) = \sum_{i=1}^{N} (x_i - \hat{x}_i)^2,
\end{equation}
where $\mathbf{x}$ is an input encoded by the encoder network, while $\hat{\mathbf{x}}$ serves as a reconstruction (the output of the decoder network).
 We employed a Multi-Layer Perceptron (MLP)  architecture to compress the $N=784$ dimensional pixel space into a highly compressed latent space of $N_{AE}$ dimensions. The value of $N_{AE}$ depends on the training dataset, see Table~\ref{tab1} ($N=N_{AE}$ for AE-NLM).  The encoder consists of a hidden layer with $400$ units. The decoder mirrors this structure, mapping the  $N_{AE}$-dimensional latent vector back to $784$ dimensions via a $400$-unit hidden layer, with a $\tanh$ activation.
 
 The AE was trained for 50 epochs using the Adam optimizer with a learning rate of $1 \times 10^{-3}$ and a batch size of 128. Following training, the entire dataset of $N_{data}=10\,000$ samples was passed through the encoder, producing a static latent dataset matrix of shape $N_{AE} \times 10\,000$, upon which the NLM was subsequently trained.
A detailed algorithm is given in Alg.~\ref{alg:pcd_langevin_training_latent}.

\begin{algorithm}
\caption{Training Protocol of the Latent Space Energy-Based Model or AE-NLM}
\label{alg:pcd_langevin_training_latent}
\begin{algorithmic}[1]

\State \textbf{Hyperparameters:} $N_{AE}$: latent space dimension; $g$: initial coupling strength; $T$: temperature; $\Delta t$: time increment; $n$: number of epochs; $k$: dynamic steps per update; $\eta$: learning rate; $\lambda$: weight decay parameter; $m$: mini-batch size; $M$: number of persistent chains.

\State \textbf{Initialization:}
\State Load MNIST dataset according to desired types of digits.
\State Load pre-trained deterministic encoder of AE, i.e., $E_{\theta}$.
\State Preprocess images to $[-1, 1]$ and extract latent representations: $\boldsymbol{\mu}_{data} = E_{\theta}({\bf{X}}_{scaled})$, where a rate-to-current transformation is applied (see Alg.~\ref{alg:pcd_langevin_training}).
\State Initialize coupling matrix $\matr{J} \in \mathbb{R}^{N_{AE} \times N_{AE}}$ from $\mathcal{N}(0, g^2/N_{AE})$, and strictly set diagonal $J_{ii} = 0$.
\State Initialize bias vector $\mathbf{b} \in \mathbb{R}^{N_{AE}}$ as zeros.
\State Initialize persistent latent states $\boldsymbol{\mu}_{model} \in \mathbb{R}^{N_{AE} \times M}$ drawn from $\mathcal{N}(0, \mathbf{I})$.

\State \textbf{Training Loop:}
\For{epoch $= 1$ to $n$}
    \For{each latent batch $\boldsymbol{\mu}_{batch} \in \mathbb{R}^{N_{AE} \times m}$ from DataLoader($\boldsymbol{\mu}_{data}$)}
        \State \textbf{Data Phase Gradients:}
        \State Compute local field: $\matr{H}_{batch} = \matr{J} \tanh(\boldsymbol{\mu}_{batch}) + \mathbf{b} \mathbf{1}_{m}^\top$
        \State $\nabla_{\matr{J}}^{data} = \frac{1}{T \cdot m} (-\boldsymbol{\mu}_{batch} + \matr{H}_{batch}) \tanh(\boldsymbol{\mu}_{batch})^\top$
        \State $\nabla_{\mathbf{b}}^{data} = \frac{1}{T \cdot m} (-\boldsymbol{\mu}_{batch} + \matr{H}_{batch}) \mathbf{1}_{m}$
        
        \State \textbf{Model Phase (Langevin Dynamics):}
        \State $\boldsymbol{\mu}_0 \leftarrow \boldsymbol{\mu}_{model}$
        \For{$s = 1$ to $k$}
            \State $\matr{H}_{s-1} = \matr{J} \tanh(\boldsymbol{\mu}_{s-1}) + \mathbf{b} \mathbf{1}_{M}^\top$
            \State $\boldsymbol{\delta J}_{s-1} = \matr{J}^\top (\matr{H}_{s-1} - \boldsymbol{\mu}_{s-1})$
            \State $\phi'(\boldsymbol{\mu}_{s-1}) = 1 - \tanh^2(\boldsymbol{\mu}_{s-1})$ \Comment{Element-wise}
            \State Draw noise $\boldsymbol{\mathcal{E}}_{s-1} \sim \mathcal{N}(0, \mathbf{I})$
            \State $\boldsymbol{\mu}_{s} = (1 - \Delta t) \boldsymbol{\mu}_{s-1} + \Delta t \cdot \matr{H}_{s-1} - \Delta t \cdot (\phi'(\boldsymbol{\mu}_{s-1}) \odot \boldsymbol{\delta J}_{s-1}) + \sqrt{2T \Delta t} \, \boldsymbol{\mathcal{E}}_{s-1}$
        \EndFor
        \State Update persistent states $\boldsymbol{\mu}_{model} \leftarrow \boldsymbol{\mu}_{k}$.
        
        \State \textbf{Model Phase Gradients:}
        \State $\matr{H}_{model} = \matr{J} \tanh(\boldsymbol{\mu}_{model}) + \mathbf{b} \mathbf{1}_{M}^\top$
        \State $\nabla_{\matr{J}}^{model} = \frac{1}{T \cdot M} (-\boldsymbol{\mu}_{model} + \matr{H}_{model}) \tanh(\boldsymbol{\mu}_{model})^\top$
        \State $\nabla_{\mathbf{b}}^{model} = \frac{1}{T \cdot M} (-\boldsymbol{\mu}_{model} + \matr{H}_{model}) \mathbf{1}_{M}$
        
        \State \textbf{Parameter Update (Gradient Descent):}
        \State Set diagonal $\nabla_{J_{ii}}^{data} - \nabla_{J_{ii}}^{model} = 0$
        \State $\matr{J} \leftarrow \matr{J} - \eta (\nabla_{\matr{J}}^{data} - \nabla_{\matr{J}}^{model} + \lambda \matr{J})$
        \State $\mathbf{b} \leftarrow \mathbf{b} - \eta (\nabla_{\mathbf{b}}^{data} - \nabla_{\mathbf{b}}^{model})$ \Comment{No weight decay for bias}
    \EndFor 
\State \textbf{Store} the trained parameters $\mathbf{J}_{\text{trained}}$ and $\mathbf{b}_{\text{trained}}$ if necessary.
\EndFor 
\end{algorithmic}
\end{algorithm}

\subsection{Generation process}
During the generation of new images, the simulation starts from a batch of initial states $\matr{X}_0\in\mathbb{R}^{N\times N_{gen}}$, each column of which is
 drawn independently from an isotropic Gaussian distribution $\mathcal{N}(\mathbf{0}, \id_N)$. $N_{gen}$ denotes the total number of generated images.
 The Langevin dynamics are simulated numerically using the same Euler-Maruyama discretization scheme as employed during the model phase of training [derived from Eq.~\ref{eq:langevin_generation} in the main text]:
\begin{equation}
    \mathbf{X}_{t+dt} = \mathbf{X}_t + dt \cdot {\bf{F}}(\mathbf{X}_t | \matr{J}_{\text{final}},\mathbf{b}) + \sqrt{2T dt} \boldsymbol{\mathcal{E}}_t,
    \label{eq:langevin_discrete_generation}
\end{equation}
where ${\bf{F}}(\mathbf{X}_t | \matr{J}_{\text{final}}) = -\nabla_{\mathbf{X}} E(\mathbf{X}_t | \matr{J}_{\text{final}})$ is the deterministic force vector evaluated with the trained parameters. The component-wise expression for the force $F_i$ is given by Eq.~\ref{eq:force_comp} in the main text. When $\mathbf{X}_t$ is a matrix representing a batch of states (e.g., $\mathbf{X}_t \in \mathbb{R}^{N \times N_{gen}}$), the update rule can be written in matrix form as:
\begin{equation}
 \matr{X}_{t+dt} = (1 - dt) \matr{X}_t + dt \cdot (\mathbf{b}\mathbf{1}^\top+\matr{J}_{\text{final}} \tanh(\matr{X}_t)) - dt \cdot (\phi'(\matr{X}_t) \odot (\matr{J}_{\text{final}}^\top (\matr{J}_{\text{final}} \tanh(\matr{X}_t)+\mathbf{b}\mathbf{1}^\top - \matr{X}_t))) + \sqrt{2T \, dt} \, \boldsymbol{\mathcal{E}}_t, 
\end{equation}
where $\phi'(x) = 1 - \tanh^2(x)$ is the derivative of $\phi(x)=\tanh(x)$, $\odot$ denotes the Hadamard (element-wise) product, and $\boldsymbol{\mathcal{E}}_t$ is a matrix of i.i.d. standard Gaussian random variables of the same dimension as $\matr{X}_t$. This simulation is run for a specified number of steps, $t_G$, until a generated image is obtained.
The generation process is outlined in Alg.
~\ref{alg:generation_process}. Hyperparameters used in all figures (including those in the main text) are summarized in Table~\ref{tab1}, while Table~\ref{tab2} summarizes the description for each hyperparameter. Codes are available in our GitHub~\cite{code-2025}.

\begin{algorithm}
\caption{Generation Process of NLM}
\label{alg:generation_process}
\begin{algorithmic}[1]

\State \textbf{Hyperparameters:} $N$: network size; $T$: temperature; $\Delta t$: time increment; $t_{G}$: total number of generation steps; $N_{gen}$: number of generated samples.

\State \textbf{Initialization:} 
\State Load the previously trained parameters $\matr{J}_{\text{final}}$ and $\mathbf{b}_{\text{final}}$. 
\State Initialize states ${\bf{X}}_0 \in \mathbb{R}^{N \times N_{gen}}$ drawn independently from standard normal noise $\mathcal{N}(\mathbf{0}, \mathbf{I})$.

\State \textbf{Generation Loop:}
\State Set $\matr{J} \leftarrow \matr{J}_{\text{final}}$ and $\mathbf{b} \leftarrow \mathbf{b}_{\text{final}}$.
\For{$s = 1$ to $t_{G}$}
    \State $\matr{H}_{s-1} = \matr{J} \tanh({\bf{X}}_{s-1}) + \mathbf{b} \mathbf{1}_{N_{gen}}^\top$ \Comment{Compute local field with bias}
    \State $\boldsymbol{\delta J}_{s-1} = \matr{J}^\top (\matr{H}_{s-1} - {\bf{X}}_{s-1})$
    \State $\phi'({\bf{X}}_{s-1}) = 1 - \tanh^2({\bf{X}}_{s-1})$
    \State Draw noise $\boldsymbol{\mathcal{E}}_{s-1} \sim \mathcal{N}(\mathbf{0}, \mathbf{I})$
    \State ${\bf{X}}_{s} = (1 - \Delta t) {\bf{X}}_{s-1} + \Delta t \cdot \matr{H}_{s-1} - \Delta t \cdot (\phi'({\bf{X}}_{s-1}) \odot \boldsymbol{\delta J}_{s-1}) + \sqrt{2T \Delta t} \, \boldsymbol{\mathcal{E}}_{s-1}$
\EndFor
\State Let ${\bf{X}}_{\text{final}} \leftarrow {\bf{X}}_{t_{G}}$
\State \textbf{Return} the generated states mapped back to visual pixel space $\tanh({\bf{X}}_{\text{final}})$.
\end{algorithmic}
\end{algorithm}

\begin{table}[H]
\centering
\caption{Hyperparameters for figures in the main text and appendix}\label{tab1}
\begin{ruledtabular}
\begin{tabular}{ccccccccccccccc}
%\hline
Figure &method  & $g$   & $N$ &$N_{data}$   & $dt$   & $k$ & $t_{age}$   & $\eta$   & $\lambda$   & $T$   & $m$   & $M$ & $N_{gen}$ &$t_G$   \\
\hline
1     &PCD-10 & 2   & 784 &288 & 0.01  & 10  & 20000   & $5\times10^{-5}$   & $5\times10^{-7}$   & 1   & 288 &1000   & 288  &-  \\
2     &Rdm-500 & 2   & 784 &10000  & 0.01  & 500  & -   & $5\times10^{-5}$   & $5\times10^{-7}$   & 1   & 1000 &1000   & 10000 &-    \\
3      &PCD-10 & 2   & 784 &-  & 0.01  & 10  & 20000   & $5\times10^{-5}$   & $5\times10^{-7}$   & 1   & 1000 &1000   & - &50000  \\
6 &PCD-10 &2 &784&10000  &0.01 & 10 &-  &$5\times10^{-5}$   & $5\times10^{-7}$     & 1   & 1000 &1000  & - &-  \\
7 &PCD-10 &2 &784&-  &0.01 & 10 &20000  &$5\times10^{-5}$   & $5\times10^{-7}$     & 1   & 1000 &1000  & - &-  \\
8   &AE-NLM & 2   & 30 &10000  & 0.01  & 10  & -   & $5\times10^{-5}$   & $5\times10^{-7}$     & 1   & 1000 &1000   & 10000 &-  \\
12 &PCD-10 &2 &784&10000  &0.01 & 10 &5000  &$5\times10^{-5}$   & $5\times10^{-7}$     & 1   & 1000 &1000   & 1000 &24939  \\
13 &AE-NLM &2 &20&10000  &0.01 & 10 &-  &$5\times10^{-5}$   & $5\times10^{-7}$     & 1   & 1000 &1000   & 1000 &-  \\
14 &PCD-600 &2 &784&1000  &0.01 & 600 &-  &$5\times10^{-4}$   & $5\times10^{-6}$     & 1   & 1 &1   & 1000 &-  \\
15 &PCD-2600 &2 &784&10000  &0.01 & 2600 &-  &$5\times10^{-4}$   & $5\times10^{-6}$     & 1   & 100 &1   & 1000 &-  \\
%\hline
\end{tabular}
\end{ruledtabular}
\end{table}

\begin{table}[H]
\centering
\caption{Description of hyperparameters used in table~\ref{tab1}}\label{tab2}
\begin{tabular}{cl}
\hline
\textbf{Symbol} & \textbf{Description} \\
\hline
$Method$ & Training method (e.g., PCD or Rdm) \\
$N$ & Network size ($N=N_{AE}$ for AE-NLM)\\
$N_{AE}$ & Dimensionality of the compressed latent space \\
$N_{data}$ & Number of training samples (dataset size) \\
$k$ & Number of sampling steps in PCD-$k$ or Rdm-$k$ \\
$n$ & Number of training epochs \\
$\eta$ & Learning rate \\
$\lambda$ & Weight decay coefficient \\
$T$ & Temperature \\
$m$ & Batch size for data averaging \\
$M$ & Number of trajectories for model averaging \\
$N_{gen}$ & Number of generated samples \\
$g$ & Initial weights $\sim \mathcal{N}(0, \frac{g^2}{N})$ \\
$dt$ & Discretization step size of time \\
$t_{age}$ & Number of parameter updates, estimated by $n\times N_{data}/m$ \\
$t_G$ & Number of dynamics steps for image generation \\
\hline
\end{tabular}
\end{table}

In particular, for the generation process of AE-NLM, the Langevin dynamics were initialized from standard normal noise in the latent space [$\mathbf{y} \sim \mathcal{N}(\mathbf{0}, \mathbf{I}_{N_{AE}})$]. The neural dynamics are carried out within the $N_{AE}$-dimensional latent space. To capture the generative trajectory, the latent states were periodically recorded at 22 logarithmic time steps defined by  $t_{G} = \text{round}(5 \times 1.5^i)$ up to a maximum of $24\,939$ steps. At each designated checkpoint, the latent states $\mathbf{y}_t$ were passed through the pre-trained AE decoder in the inference mode to project the generated trajectories back into the original $784$-dimensional pixel space for visualization and quantitative analysis. Pseudocode refers to Alg.~\ref{alg:generation_process_latent}.

\begin{algorithm}
\caption{Generation Process via Latent Space Langevin Dynamics}
\label{alg:generation_process_latent}
\begin{algorithmic}[1]

\State \textbf{Hyperparameters:} $N_{AE}$: latent space dimension; $T$: temperature; $\Delta t$: time increment; $t_{G}$: total number of generation steps; $N_{gen}$: number of generated samples.

\State \textbf{Initialization:} 
\State Load the trained NLM parameters $\matr{J}_{\text{final}}$ and $\mathbf{b}_{\text{final}}$. 
\State Load the decoder network $D_{\phi}$ of the pre-trained Autoencoder (AE).
\State Initialize latent neural states $\boldsymbol{\mu}_0 \in \mathbb{R}^{N_{AE} \times N_{gen}}$ drawn independently from standard normal noise $\mathcal{N}(\mathbf{0}, \mathbf{I})$.

\State \textbf{Generation Loop (Latent Space):}
\State Set $\matr{J} \leftarrow \matr{J}_{\text{final}}$ and $\mathbf{b} \leftarrow \mathbf{b}_{\text{final}}$.
\For{$s = 1$ to $t_{G}$}
    \State $\matr{H}_{s-1} = \matr{J} \tanh(\boldsymbol{\mu}_{s-1}) + \mathbf{b} \mathbf{1}_{N_{gen}}^\top$ \Comment{Compute local field}
    \State $\boldsymbol{\delta J}_{s-1} = \matr{J}^\top (\matr{H}_{s-1} - \boldsymbol{\mu}_{s-1})$
    \State $\phi'(\boldsymbol{\mu}_{s-1}) = 1 - \tanh^2(\boldsymbol{\mu}_{s-1})$
    \State Draw noise $\boldsymbol{\mathcal{E}}_{s-1} \sim \mathcal{N}(\mathbf{0}, \mathbf{I})$
    \State $\boldsymbol{\mu}_{s} = (1 - \Delta t) \boldsymbol{\mu}_{s-1} + \Delta t \cdot \matr{H}_{s-1} - \Delta t \cdot (\phi'(\boldsymbol{\mu}_{s-1}) \odot \boldsymbol{\delta J}_{s-1}) + \sqrt{2T \Delta t} \, \boldsymbol{\mathcal{E}}_{s-1}$
\EndFor
\State Let final latent states $\boldsymbol{\mu '} \leftarrow \boldsymbol{\mu}_{t_{G}}$.

\State \textbf{Decoding Phase:}
\State Map generated latent states back to visual pixel space: ${\bf{X}}_{\text{gen}} = D_{\phi}(\tanh(\boldsymbol{\mu '} ))$, where a current-to-rate transformation is applied.
\State \textbf{Return} the generated images ${\bf{X}}_{\text{gen}}$.
\end{algorithmic}
\end{algorithm}

\section{Evaluation of generation quality via adversarial accuracy indicator}\label{SM-c}
To assess the fidelity of the generated samples at different values of training age ($t_{age}$), we evaluated the generative performance of NLM and its variants using a sample size of $N_s$ real images (source) and $N_t$ generated images (target). \blue{Note that the sizes of source and target are the same in this paper.} We then define $Acc_T$ as the probability that a generated sample has
a nearest neighbor (n.n.) which is also a generated sample, while $Acc_S$ is defined as the probability that a data sample has a n.n. which is
also a data sample.  If both probabilities achieve $0.5$, the real and generated data samples are perfectly mixed. Therefore, a natural evaluation metrics is given below.
\begin{equation}
    {\cal E}^{{\rm{AAI}}} = \frac{1}{2} \left[ (Acc_S - 0.5)^2 + (Acc_T - 0.5)^2 \right]
\end{equation}
where AAI is a shorthand notation of adversarial accuracy indicator.

Analyzing $Acc_S$ and $Acc_T$ separately reveals three critical limits of the generative behavior (Fig.~\ref{figure11}), as shown below.
\begin{itemize}
    \item \textbf{Complete Separation ($Acc_S \approx 1, Acc_T \approx 1$):} The generated distribution and the real distribution are disjoint in the data space, resulting in a maximum of ${\cal E}^{{\rm{AAI}}}$.
    \item \textbf{Perfect Mixture ($Acc_S \approx 0.5, Acc_T \approx 0.5$):} This represents the ideal generative performance. Real and generated samples are indistinguishable, thereby minimizing ${\cal E}^{{\rm{AAI}}}$ to zero.
    \item \textbf{Overfitting/Memorization ($Acc_S \approx 0, Acc_T \approx 0$):} The model perfectly memorizes the training data. 
    \end{itemize}

\begin{figure}[htbp]
    \centering
    \includegraphics[width=0.7\textwidth]{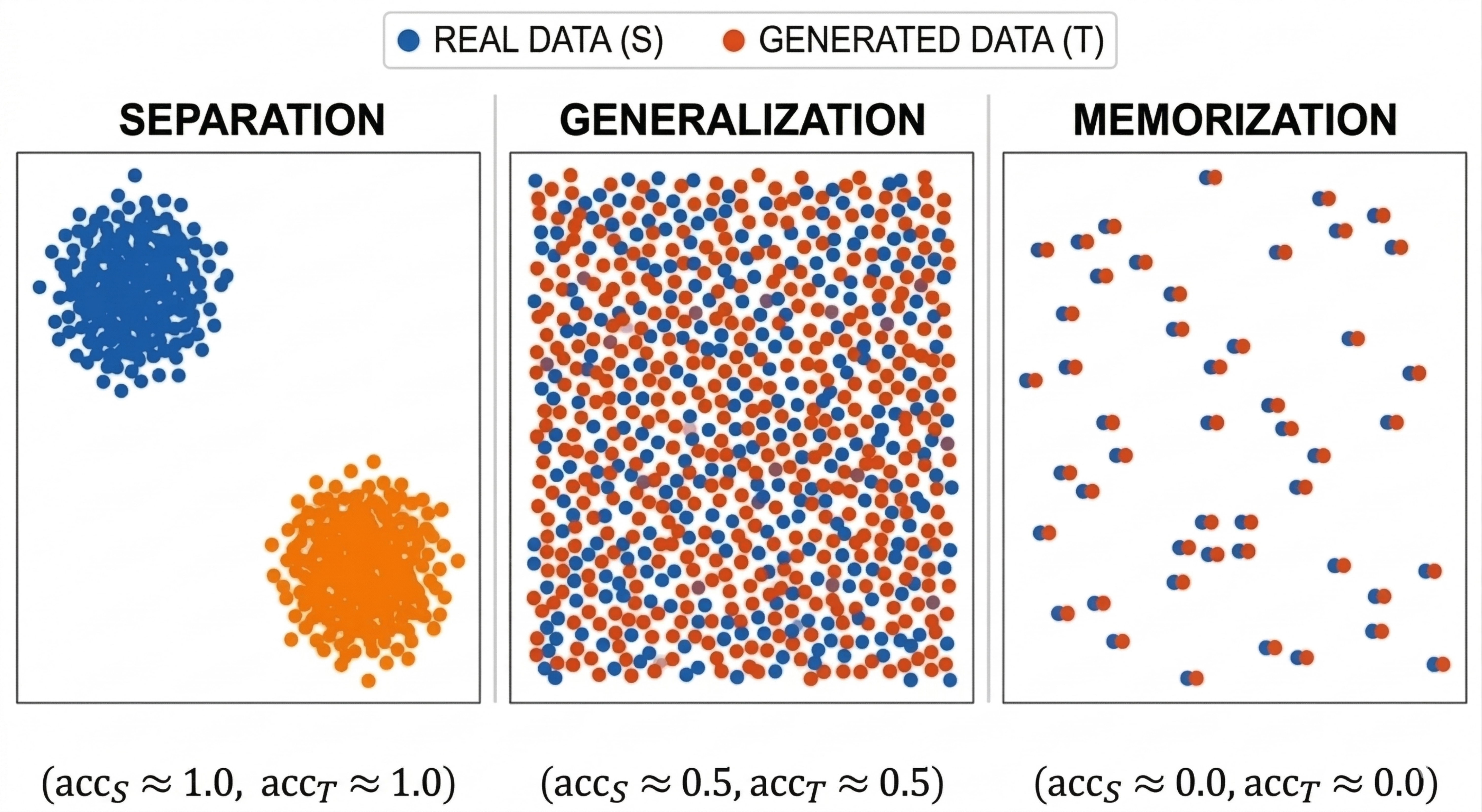}
    \caption{Visual interpretation of adversarial accuracy indicator. \textbf{Separation} ($acc_S$ or $acc_T\approx 1.0$): the model failed to capture the data distribution; \textbf{Generalization} ($acc_S$ and $acc_T \approx 0.5$): real and generated data samples are indistinguishable; \textbf{Memorization} ($acc_S$ or $acc_T \approx 0.0$): the overfitting leads to memorization of training samples.}
    \label{figure11}
\end{figure}

\section{Other results on performance of NLM}\label{SM-d}
\begin{figure}[htbp]
\centering
\includegraphics[width=0.5\textwidth]{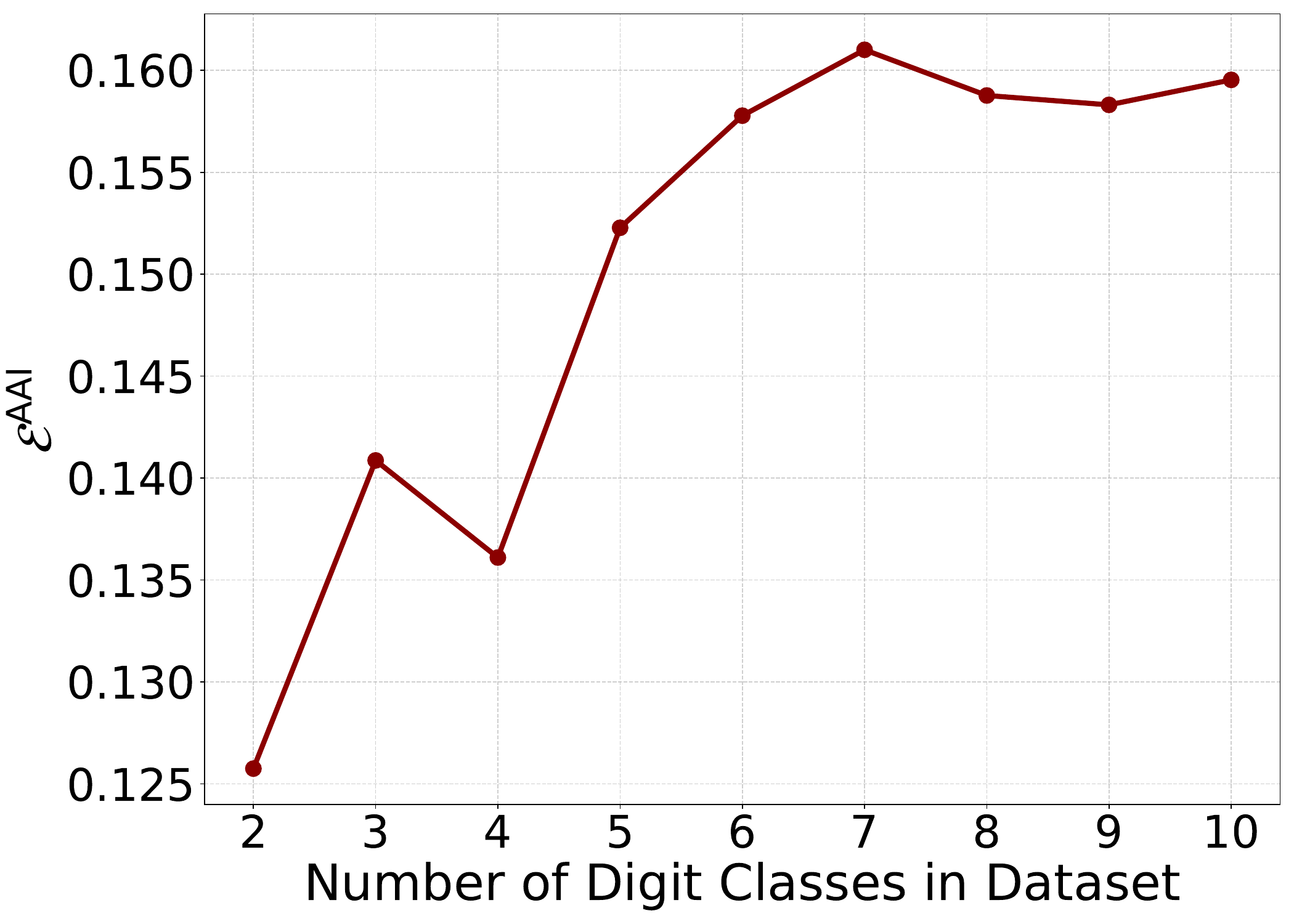}
\caption{$\mathcal{E}^{\text{AAI}}$ versus the number of digit classes. $N_{\rm data}=10000$. PCD-10 is used.}
\label{figure12}
\end{figure}

\begin{figure}[htbp]
\centering
\includegraphics[width=0.9\textwidth]{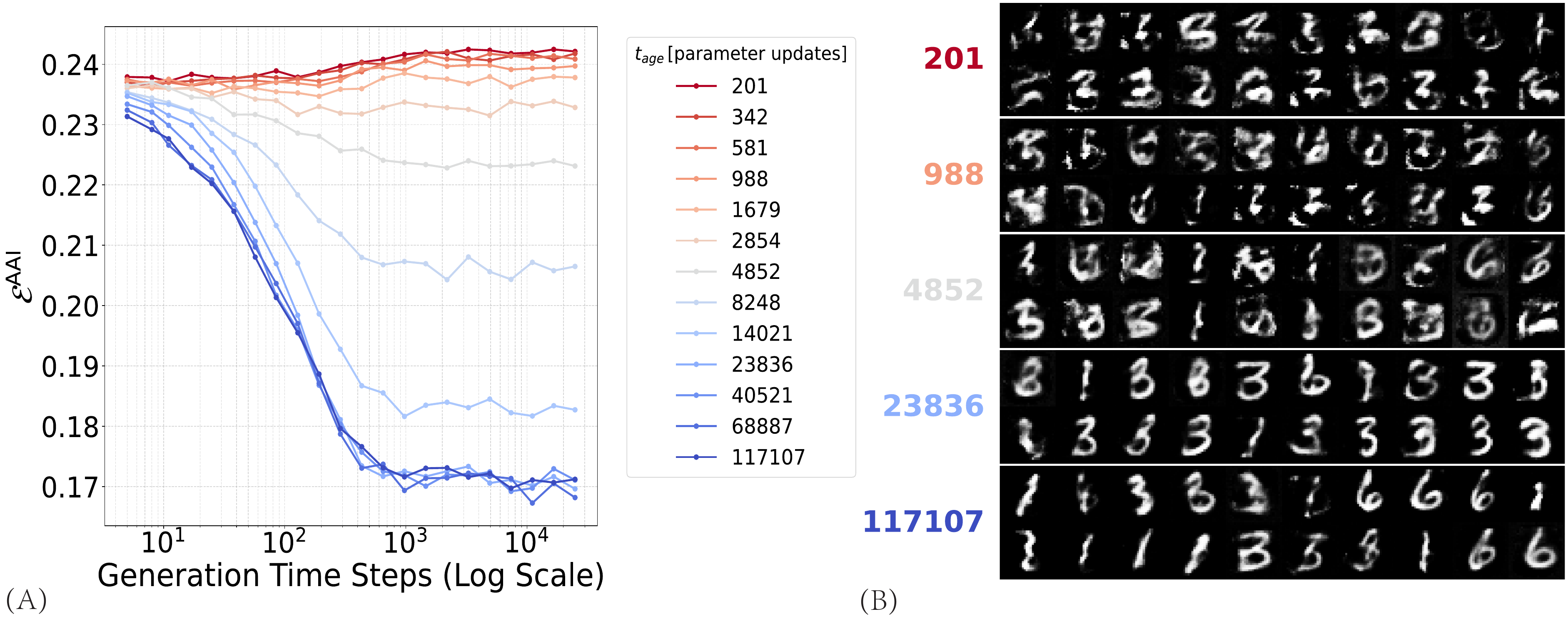}
\caption{Performance of AE-NLM. (A) $\mathcal{E}^{\text{AAI}}$ versus $t_G$. (B) Generated samples at different values of $t_{\rm age}$. }
\label{figure13}
\end{figure}

The performance of NLM trained using PCD-10 degrades as the number of digit class increases, as shown in Fig.~\ref{figure12}. However, the quality can be enhanced by applying AE-NLM, whose performance is shown in Fig.~\ref{figure13}.
When applied to more complex dataset Fashion-MNIST, the quality is also enhanced (see Fig.~\ref{fig8} in the main text) \textit{even if} considering all types of images. 

\begin{figure}[htbp]
    \centering
     \includegraphics[width=0.9\textwidth]{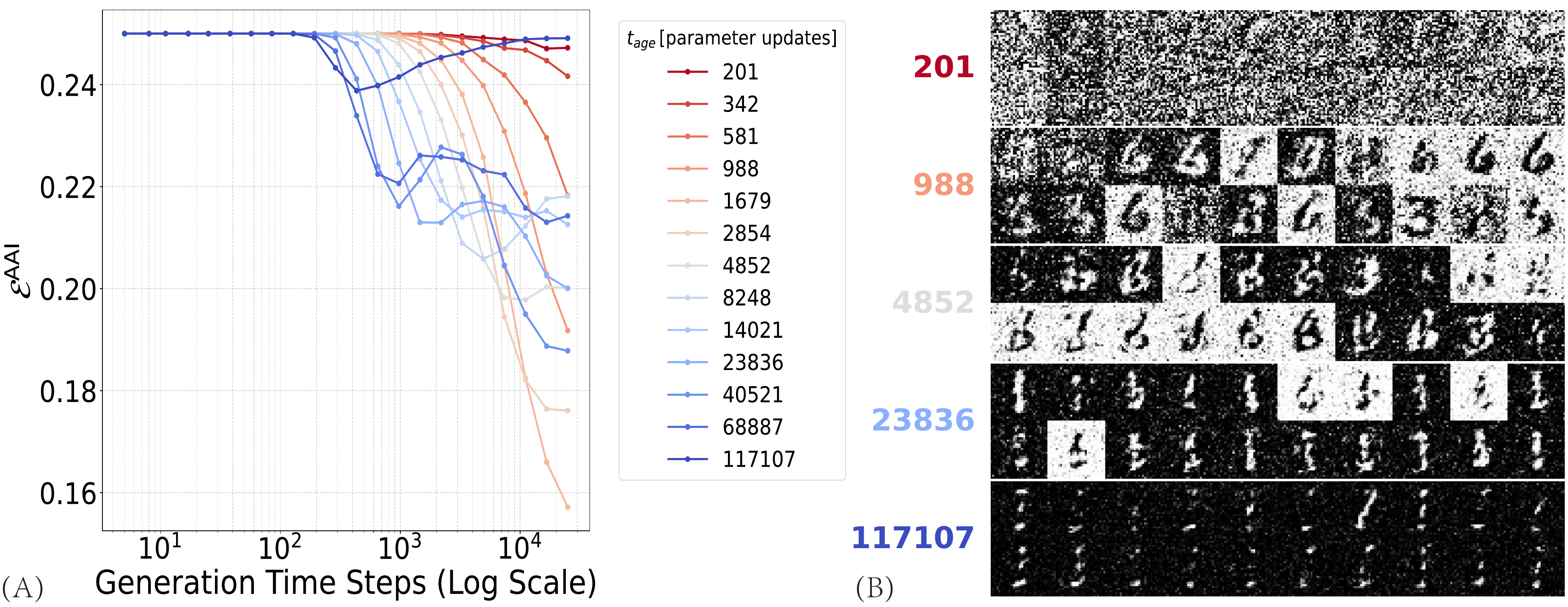}
    \caption{ Generative performance of PCD-600 trained using one sample for data and model phase (no data or model average).  \textbf{(A)} The adversarial accuracy indicator (${\cal E}^{{\rm{AAI}}}$) evaluated over generation time steps ($t_G$). The visual degradation of the over-trained model observed in (B) strictly correlates with a corresponding increase in the AAI metrics. \textbf{(B)} Final generated samples (recorded at the maximum generation time step, $t = 24939$) produced by models at five distinct training stages (ages). Models at intermediate parameter update steps (e.g., $t_{age} = 988$ or $1679$ or $2854$) generate coherent digit structures despite a bit low quality. }
    \label{figure14}
\end{figure}

\begin{figure}[htbp]
    \centering
     \includegraphics[width=0.9\textwidth]{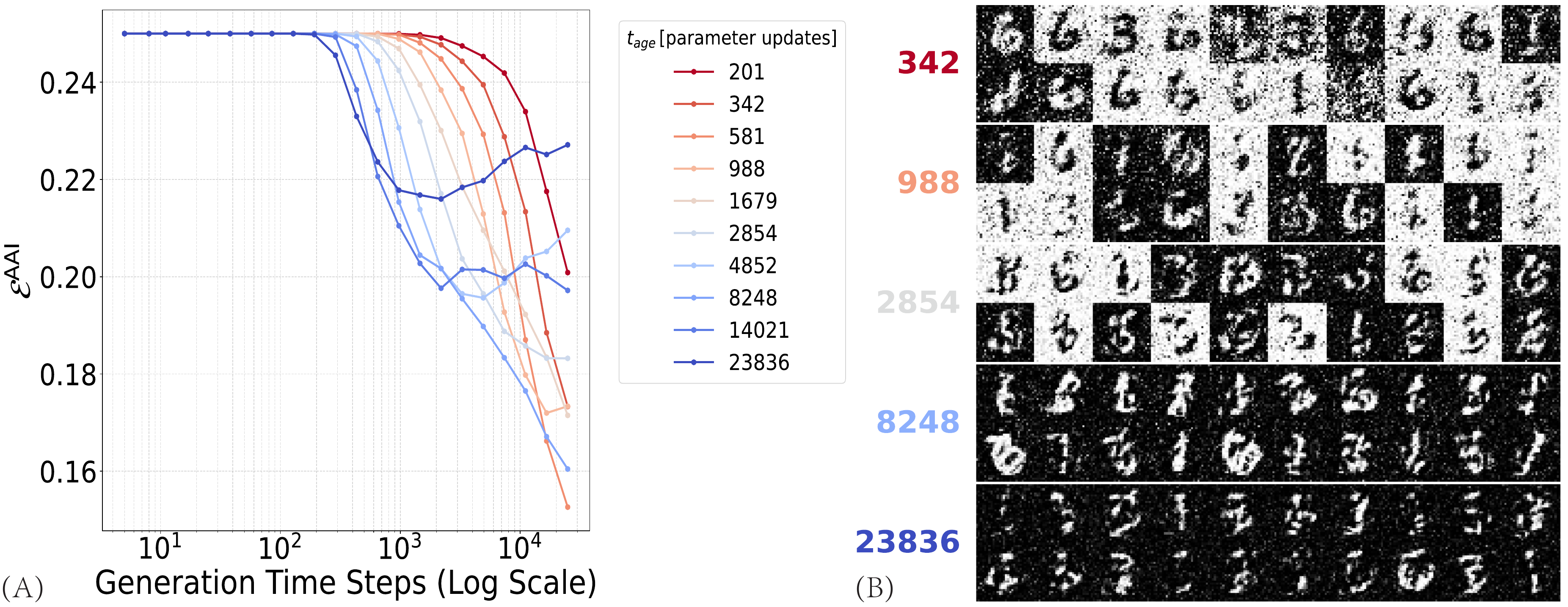}
    \caption{ Generative performance of PCD-2600 trained using one persistent chain, where every 2600 dynamics step a parameter update is carried out and 100 samples (collected every 20 steps) are accumulated for model average.  \textbf{(A)} The adversarial accuracy indicator (${\cal E}^{{\rm{AAI}}}$) metrics evaluated over generation time steps ($t_G$). The visual degradation of the over-trained model observed in (B) strictly correlates with a corresponding increase in the AAI metrics. \textbf{(B)} Final generated samples (recorded at the maximum generation time step, $t = 24939$) produced by models at five distinct training stages (ages).  }
    \label{figure15}
\end{figure}

To avoid running many parallel chains to collect samples for model average, we also test one chain for one sample and find that NLM still works as shown in Fig.~\ref{figure14}.  The results of one-chain many samples (sampled with fixed interval, accumulated for model average or a direct exponential averaging, i.e., $\mu\leftarrow(1-\alpha)\mu+\alpha f(x)$, where $\alpha$ is a weight coefficient and $f(x)$ indicates the relevant function relying on the current sample) are also shown in Fig.~\ref{figure15}, which demonstrates the feasibility of one persistent chain sampling (resembling a biologically neural dynamics sampling).

\section{NLM as an associative memory network}\label{SM-e}
In the original recurrent neural network~\cite{Chaos-1988}, it is verified that an exponential number of unstable fixed points exist in the phase space~\cite{Helias-2025}. In the main text,
these fixed points are reachable when the kinetic energy is optimized, yielding neural Langevin dynamics. According to this principle, we can also design an associative memory network with asymmetric couplings, where a noisy pattern can be denoised via minimizing the kinetic energy, i.e., when a fixed point is reached, a clean pattern (such as an image) is recovered. Therefore, we can train the network to store a set of images and further test the denoising ability.

We used the MNIST dataset for our experiments. To form the memory patterns, we selected 10 distinct samples, one for each digit from 0 to 9. Each $28 \times 28$ grayscale image was flattened into a 1D vector of dimension $N = 784$. The pixel values were linearly scaled from the standard $[0, 1]$ range to $[-1, 1]$ to match the domain of the $\tanh(\cdot)$ activation function. The model consists of two trainable parameters: the interaction matrix $\mathbf{J} \in \mathbb{R}^{784 \times 784}$ and the bias vector $\mathbf{b} \in \mathbb{R}^{784}$. The initialization scheme is listed below:
\begin{itemize}
    \item The elements of matrix $\mathbf{J}$ were initialized randomly from a normal distribution ${\cal N}(0,\frac{2}{{{N^2}}})$. To enforce the strict topological constraint that neurons do not self-interact, we applied a diagonal mask to ensure $J_{ii} = 0$ during every step.
    \item The bias vector $\mathbf{b}$ was initialized to zeros.
\end{itemize}

The objective of the training was to minimize the total energy function $E(\mathbf{x})$ of the system. The training details are listed below:
\begin{itemize}
    \item \textbf{Optimizer:} We utilized the Adam optimizer for gradient descent.
    \item \textbf{Learning Rate:} The learning rate was set to $0.001$.
    \item \textbf{Epochs:} The model was trained for a total of 1000 epochs.
\end{itemize}
During each training step, the parameters $\mathbf{J}$ and $\mathbf{b}$ were updated to minimize the energy $E(\mathbf{x})$. Details are given in 
the pseudocode Alg.~\ref{hopf}. 

To evaluate the model's pattern retrieval and denoising capabilities, we simulated the system dynamics using the Euler method.
\begin{itemize}
    \item \textbf{Noise Injection:} We selected one stored digit and corrupted it with $25\%$ noise. This means $25\%$ of the image pixels were randomly replaced with values uniformly sampled from the $[-1, 1]$ range. The pixels of the corrupted image are then transformed to the current domain. 
    \item \textbf{Denoising Process:} The noisy input was used to initialize the deterministic dynamics ($T=0$), which are updated iteratively using the deterministic driving force $F_i$. We ran the dynamics for 451 continuous steps using a step size of $dt = 0.01$.
\end{itemize}
Throughout the dynamic evolution, we recorded the decrease in the system's kinetic energy and extracted the intermediate images at a specific time (steps 0, 150, 300, and 450) to illustrate the recovery process, as shown in Fig.~\ref{fig10} (see the main text).

\begin{algorithm}
\caption{Training Process of NLM as an associative memory model}
\label{hopf}
\begin{algorithmic}[1]
\Require MNIST dataset, Adam optimizer, Learning rate $\eta = 0.001$, Total epochs $n = 1000$
\State \textbf{Data Preparation}
\State Select 10 distinct images $\mathbf{X} = \{\mathbf{x}^{(k)}\}_{k=1}^{10}$ (one for each digit 0-9)
\State Flatten each $28 \times 28$ image into a 1D vector of dimension $N = 784$
\State Data preprocessing: ${\bf{X}}_{data} = \text{atanh}\left(\text{clip}(2 \cdot {\bf{X}}_{raw} - 1, -1+\epsilon, 1-\epsilon)\right)$
\State \textbf{ Model Initialization}
\State Initialize interaction matrix $\mathbf{J} \sim \mathcal{N}(0, \frac{4}{N})$
\State Initialize bias vector $\mathbf{b} \leftarrow \mathbf{0}$
\State Define diagonal mask $\mathbf{M}$ where $M_{ii} = 0$ and $M_{ij} = 1$ for $i \neq j$
\State \textbf{ Training Process}
\For{$epoch = 1$ \textbf{to} $n$}
    \State Apply mask to interaction matrix: $\tilde{\mathbf{J}} = \mathbf{J} \odot \mathbf{M}$
    \State Compute local fields for all samples: $h_i^{(k)} = \sum_{j=1}^N \tilde{J}_{ij} \tanh(x_j^{(k)})+b_i$ for $k \in \{1,\dots,10\}$
    \State Compute total system energy: $E(\mathbf{X}) = \frac{1}{2}\sum_{k=1}^{10}\sum_{i=1}^N \left( -x_i^{(k)} + h_i^{(k)}  \right)^2$
    \State Compute gradients over the entire set: $\nabla_{\mathbf{J}} E(\mathbf{X})$ and $\nabla_{\mathbf{b}} E(\mathbf{X})$
    \State Update $\mathbf{J}$ and $\mathbf{b}$ using Adam optimizer with learning rate $\eta$
\EndFor
\State Apply mask to final weights: $\mathbf{J} \leftarrow \mathbf{J} \odot \mathbf{M}$
\State \textbf{Return} Trained model parameters $\mathbf{J}$ and $\mathbf{b}$
\end{algorithmic}
\end{algorithm}

%\end{widetext}

%%%%%%%%%%%%%%%%%%%%%%%%%%%%%%%%%%%%%%%%%%%%%%%%%%%%%%%%%%%%%%%%%%%%%
%\bibliography{ref}

%%%%%%%%%%%%%%%%%%%%%%%%%%%%%%%%%%%%%%
\end{document}